\documentclass[iop,apj]{emulateapj}

\usepackage{apjfonts}

\usepackage{amsmath}
\usepackage{graphicx}
\usepackage{epstopdf}
\usepackage{epsfig}
\usepackage{natbib}
\usepackage{wasysym}
\usepackage{lipsum}
\usepackage{mathrsfs}
\usepackage{float}
\usepackage{rotating}
\usepackage{amsmath} % or simply amstext
\newcommand{\angstrom}{\textup{\AA}}
\providecommand{\e}[1]{\ensuremath{\times 10^{#1}}}

\usepackage{color,hyperref}

\usepackage{color}

\shorttitle{Evolving Gaseous Sub-Neptunes with MESA}
\shortauthors{Chen \& Rogers}

\begin{document}

\title{Evolutionary Analysis of Gaseous Sub-Neptune-Mass Planets with MESA}

\author{Howard Chen$^{1}$ \& Leslie A. Rogers$^{2,3,4,5,6}$} \footnotetext[5]{Hubble Fellow}
\footnotetext[6]{NASA Sagan Fellow}

\affil{$^1$Department of Physics, Boston University, 590 Commonwealth Ave., Boston, MA 02215, USA} 
\affil{$^2$Department of Astronomy, California Institute of Technology, MC249-17, 1200 East California Boulevard, Pasadena, CA 91125, USA}
\affil{$^3$Division of Geological and Planetary Sciences, California Institute of Technology, MC249-17, 1200 East California Boulevard, Pasadena, CA 91125, USA}
\affil{$^4$Department of Earth \& Planetary Sciences, University of California, 307 McCone Hall, Berkeley, CA, 94720-4767, USA}

\email{howardchen2021@u.northwestern.edu}

\begin{abstract}
    Sub-Neptune-sized exoplanets represent the most common types of planets in the Milky Way, yet many of their properties are unknown. Here, we present a prescription to adapt the capabilities of the stellar evolution toolkit Modules for Experiments in Stellar Astrophysics (MESA) to model sub-Neptune mass planets with H/He envelopes. With the addition of routines treating the planet core luminosity, heavy element enrichment, atmospheric boundary condition, and mass loss due to hydrodynamic winds, the evolutionary pathways of planets with diverse starting conditions are more accurately constrained. Using these dynamical models, we construct mass-composition relationships of planets from 1 to 400 $M_{\oplus}$ and investigate how mass-loss impacts their composition and evolution history. We demonstrate that planet radii are typically insensitive to the evolution pathway that brought the planet to its instantaneous mass, composition and age, with variations from hysteresis $\la 2\%$. We find that planet envelope mass loss timescales, $\tau_{\rm env}$, vary non-monotonically with H/He envelope mass fractions (at fixed planet mass). In our simulations of young (100~Myr) low-mass ($M_p\lesssim10~M_\oplus$) planets with rocky cores, $\tau_{\rm env}$ is maximized at $M_{\rm env}/M_p=1\%$ to 3\%. The resulting convergent mass loss evolution could potentially imprint itself on the close-in planet population as a preferred H/He mass fraction of ${\sim}1\%$. Looking ahead, we anticipate that this numerical code will see widespread applications complementing both 3-D models and observational exoplanet surveys.
\end{abstract}


\section{Introduction} 
\label{sec:intro} 

A striking revelation from NASA's {\it Kepler} mission is the profusion of sub-Neptune sized planets discovered with short orbital periods ($P_{\rm orb} \la 50$ days) \citep{BoruckiEt2011ApJ,BatalhaEt2013ApJ, BurkeEt2014ApJS,RoweEt2014ApJ,HanEt2014PASP,MullallyEt2015ApJS, BurkeEt2015ApJ}. 
Despite the absence of sub-Neptune, super-Earth sized planets in our Solar System, they are quite ubiquitous in the Milky Way, comprising the majority of planets found by {\it Kepler} \citep{FressinEt2013ApJ, PetiguraEt2013ApJ}.  For inner orbital distances $\lesssim 0.25$~AU, the number of sub-Neptune mass planets exceeds that of Jovian planets by more than a factor of 30 \citep{HowardEt2012ApJ}. 

Among the transiting sub-Neptune-sized planets that have measured masses, many of them have mean densities so low that they must have significant complement of light gasses (hydrogen and helium) contributing to the planet volume \citep[e.g., Kepler-11c,d,e,f,g][]{LissauerEt2011Nature, LissauerEt2013ApJ}. \citet{Wolfgang+Lopez2015ApJ} inferred from the {\it Kepler} radius distribution that most sub-Neptunes should have present day composition of ${\sim}1\%$ H/He envelope, under the assumption that all close-in planets consist of rocky cores surrounded by H/He envelopes.  Considering the sample of {\it Kepler} transiting planets with Keck-HIRES radial velocity follow-up \citep{MarcyEt2014ApJS}, \citet{Rogers2015ApJ} showed that at planet radii of $1.6~R_{\oplus}$ (and larger) most close-in planets of that size have sufficiently low mean densities that they require a volatile envelope (consisting of H/He and/or water). 
Notably, a large scattering of mass-radius measurements is found in the sub-Neptune, super-Earth size regime \citep{MarcyEt2014ApJS}, particularly between 1.6 and 4 $R_{\oplus}$ \citep{Weiss&Marcy2014ApJL}. Noise in the mass-radius measurements (which often have large error bars) does not account for all of the apparent scatter; there is evidence for intrinsic dispersion in the masses of planets of specified size \citep{WolfgangEt2015}. 
The intrinsic scatter of small planet mass-radius measurements is an indicator of compositional diversity \citep[e.g.,][]{Rogers&Seager2010ApJ}. 

The composition distribution of planets observed today reflects both the initial outcomes from planet formation, and subsequent post-formation evolution processes. 
To study the latter, a useful avenue is numerical simulations of planets' thermo-physical evolution. Stellar flux delays the cooling and contraction of close-in planets. At the same time, higher levels of incident flux also mean greater susceptibility to atmospheric escape. By "backtracking" the thermal and mass loss evolution history, inferences about the present-day composition and past history of a planet can be made. Such analyses have been carried out for GJ 1214b, CoRoT-7b, and the Kepler 11 and Kepler 36 systems \citep{ValenciaEt2010A&A, NettelmannEt2011ApJ,LopezEt2012ApJ,Lopez&Fortney2013ApJ,Howe&Burrows2015}.

The effect of atmospheric escape has also been studied for exoplanets in the broader population-level context. For example, the planet mass-loss simulations of \citet{Lopez&Fortney2013ApJ}, \citet{Owen&Wu2013ApJ}, and \citet{JinEt2014ApJ} predict a ``radius occurrence valley" dividing the sub-populations of close-in planets that have lost/retained their volatile envelopes. Focusing on higher-mass planets, \citet{Kurokawa&Nakamoto2014ApJ} investigated whether mass-loss from hot-Jupiters could reproduce the observed ``desert of sub-Jupiter size exoplanets." More recently, \citet{LugerEt2015ApJ} assessed migration, habitability, and transformation of Neptune-like to Earth-like worlds orbiting M-dwarfs.

In this work, we make two contributions. 
First, we implement several physical formulations related to low mass planets with hydrogen/helium envelopes into the state-of-the-art MESA code (Modules for Experiments in Stellar Astrophysics) \citep{PaxtonEt2011ApJ, PaxtonEt2013, PaxtonEt2015ApJS}.
MESA is an open-source, 1-D stellar evolution code that has seen wide use to address problems in stellar astrophysics such as low mass and high mass stars, white dwarfs, young neutron stars, and pre-supernova outbursts or supernova core-collapse \citep[e.g.,][]{WolfEt2013MESA,PernaEt2014ApJ,Mcley&Soker2014}. Only a handful of planetary studies using MESA exist in the literature \citep{Batygin&Stevenson2013ApJ, Owen&Wu2013ApJ, ValsecchiEt2014ApJ, ValsecchiEt2015ApJ,JacksonEt2016arXiv}, and there is still much potential to push the numerical code to even lower planet masses ($1 - 10~M_{\oplus}$). It is our aim that this paper will serve as a platform for future exoplanetary studies with MESA. The open source nature of MESA allows the astronomical community to readily access the numerical extensions introduced here. 

Second, we generate suites of planet evolution  simulations to compute mass-radius-composition-age relations for low mass planets, to quantify how those relations depend on evolution history, and, finally, to explore whether photo-evaporation can produce a ``favored" planet composition (envelope mass fraction). The numerical models and results obtained are available for public access.

This paper is structured as follows. In Section~\ref{sec:meth}, we describe methods for modeling low-mass planets with MESA. We benchmark are simulations against previously published planet evolution calculations in Section~\ref{sec:benchmark}. We present our numerical results for the coupled thermal-mass-loss evolution of planets in Section~\ref{sec:results}, and discuss and conclude in Sections~\ref{sec:discussion} and ~\ref{sec:sum}.

\section{Model \& Approach}
\label{sec:meth}

We employ the MESA toolkit \citep{PaxtonEt2011ApJ, PaxtonEt2013, PaxtonEt2015ApJS} (version 7623) to construct and evolve thousands of planet models. We consider spherically symmetric planets consisting of a heavy-element interior (comprised of rocky material, or a mixture of rock and ice) surrounded by a hydrogen-helium dominated envelope. The one-dimensional stellar evolution module, MESA star, is adapted to evolve planetary H/He envelopes.

In simulating planetary H/He envelopes, we employ the default MESA input options, unless otherwise stated. 
For equation of state in planetary conditions, we adopted the hydrogen/helium equation of state (EOS) from \citet{SaumonEt1995ApJS}. For the sake of simplicity, we restrict ourselves to solar values of metallicity $Z = 0.03$ and helium fraction $Y = 0.25$, unless otherwise stated. We used the standard low temperature Rosseland tables \citep{FreedmanEt2008} and \citep{FreedmanEt2014} for visible and infrared opacities. 
The MESA EOS and opacity tables are further  described in \citet{PaxtonEt2011ApJ, PaxtonEt2013}.

MESA already has several useful built-in functions designed for the study of planets (e.g., \citealt{PaxtonEt2013}, \citealt{Wu&Lithwick2013ApJ}, \citealt{Becker&Batygin2013ApJ}, and \citealt{ValsecchiEt2014ApJ}). Some of these build-in capabilities however, run into issues when trying to create and evolve small planets. The following sections (Section~\ref{subsec:entropy} to \ref{subsec:core}) detail our modifications to the MESA code, and recipes for constructing low-mass planets.

\subsection{Initial Starting Models}
\label{subsec:entropy}

%Overview of how to create an initial model in MESA
Creating an initial starting model in MESA is a multi-step process. First, we make an initial model (using the create\_initial\_model option) with a fixed pressure and temperature boundary. We specify these values to be comparable to those in the subsequent evolutionary stages. This technique avoids large discrepancies between the created model and parameters imposed in the evolution phase.

Next, we insert an inert core (using the relax\_core option) at the bottommost zone, to represent the heavy-element interior of the planet.  We use the models of \citet{RogersEt2011ApJ} to determine the core radius and bulk density. We consider both a "rocky composition" (70\% silicates and 30\% Fe) and an "ice-rock mixture composition" (67\% H$_{\rm 2}$O, 23\% silicates and 10\% Fe) as options for the heavy-element interior. The addition of an inert core at this stage allows the envelope to be more strongly gravitationally bound in the steps that follow. 

In the next step, we rescale the entire planet envelope to the desired total planet mass (using the relax\_mass\_scale option). Another option to reduce the planet mass is via a mass-loss wind \citep[relax\_mass][]{Batygin&Stevenson2013ApJ}.  

Finally, after having constructed a model with the desired core mass and envelope mass, we still must quantitatively standardize the initial entropy starting conditions before entering the evolution phase. In each simulation, we take the general approach of ``re-inflating" planets, to reset planet evolution.

For high mass planets $\left(M_p\geq17~M_{\oplus}\right)$, we use an artificial core luminosity to re-inflate the planet envelope until it reaches a specified interior entropy threshold (which depends on mass). In a discrete set of test cases, we determined the maximum entropy $\left(s_{\rm max}\right)$ to which the planets could be inflated before they reach runaway inflation. We fit a linear function to the maximum entropy data as a function of planet mass, $\left(s_{\rm max}/\left(\mathrm{k_B\,baryon^{-1}}\right) = 2.1662 \log\left({M_p}/{M_{\oplus}}\right) + 10.6855\right)$, and use this interpolating function to specify the initial entropy of our high-mass simulated planets. For planets with $M_p\geq17~M_{\oplus}$, the artificial core luminosity deposited at the base of the envelope during the reinflation phase is taken to be three times the surface luminosity (total intrinsic luminosity) of the planet after it is left to passively evolve for 1000 years at the end of the mass-reduction phase.

%How to reinflate low-mass planets
For low mass models $\la 17~M_{\oplus}$, special care must be taken when reinflating the planet envelopes. %Fix
In particular, the procedure used to set the artificial reinflation luminosity in the high mass regime (three times the planet luminosity at the end of the mass-reduction phase) can blow the envelope of these low-mass loosely bound planets apart. In the low-mass regime, we took a discrete set of test cases (spanning a grid of $M_p$ and $f_{\rm env}$) and determined in each case the maximum reinflation luminosity factor for which the planet envelope would remain bound (within a precision of 0.02). In practice this involved finding the highest artificial core luminosity for which the planet radius plateaued to a new equilibrium within 100~Myr, instead of expanding to infinity. Then, given any arbitrary planet mass and envelope mass fraction as inputs, we use 2D interpolation to resolve the specific luminosity.

Once the planet is reinflated, we turn off the artificial reinflation luminosity, and reset the planet age to zero. The zero-age planet model that we have created so far, does not yet account for radiation from the host star. We then use the MESA relax\_irradiation option to gradually irradiate our non-irradiated model to the desired level. Convergence issues may arise if a high irradiation flux is abruptly introduced on an unirradiated model; not only does the irradiation have to diffuse its way in from the surface of the envelope, but the heat flux escaping from the core also has to adjust to new surface boundary conditions. The relax\_irradiation option solves this issue by adding a phase in which the flux is turned on slowly.

After the initial starting conditions have been set, we allow the planet to undergo pure thermal evolution (without mass-loss) for 10~Myr before turning on mass loss \citep[following][]{LopezEt2012ApJ}. These initial cooling periods avoid the runaway mass loss scenario where the size of planet increase without bound as the simulations enter the mass-loss evolution phases. By the onset of evaporation at 10 Myr, models in general should be insensitive to the details of their initial entropy choice. This is because the cooling timescale is typically much less than 10 Myr.

\subsection{Atmospheric Boundary Conditions}
\label{subsec:atm}

%Intro paragraph
When simulating highly irradiated low-mass planets $\left(M_p\lesssim20~M_{\oplus}\right)$, direct application of MESA's default irradiated atmosphere boundary conditions can lead to problems. 
In particular, care must be taken to ensure that the atmospheric boundary conditions account for the large changes in surface gravity the low-mass low-density planets may experience over their evolution as they cool and contract.

In MESA's current grey\_irradiated planetary atmosphere option, the surface pressure (at the base of the atmosphere) is resolved at a fixed value. This can lead to issues for simulations of low-mass planets $\left(\lesssim20~M_{\oplus}\right)$ contracting from very puffy initial states, as the optical depth and opacities at a specified pressure level vary significantly over time. We follow the approach of \citet{Owen&Wu2013ApJ} to overcome this issue in implementing the $T(\tau)$ relation of \citet{Guillot2010A&A} specifying a fixed optical depth $\tau$ (instead of fixed pressure) at the base of the atmosphere. This allows the surface pressure boundary to vary over the course of the planets' evolution. Unless otherwise stated, the planet radii quoted through out this work, are defined at optical depth $\tau = 2/3$ (for outgoing thermal radiation).

To relate the irradiation flux absorbed by a planet to the planet's orbital separation and host star properties, a model for the fraction of the irradiation that is absorbed versus reflected by the planet's atmosphere is needed. In general, the Bond albedo of a planet depends on the precise atmospheric composition and the scattering properties of clouds in the planet's atmosphere. Constructing the planet atmosphere directly is beyond the scope of this article, instead we assume albedo values are taken from \citet{FortneyEt2007ApJ}. % and \citet{DemoryEt2014ApJ}.

\subsection{Hydrodynamic Evaporative Mass-Loss}
\label{subsec:massloss}

Extreme ultraviolet and X-ray radiation (EUV; $200 \lesssim \lambda \lesssim 911 \angstrom$) from a planet's host star heat the outer reaches of a planet's H/He envelope. 

For close-orbiting planets, this energy imparted on the atmosphere generates a hydrodynamic wind and causes some gas to escape the planets' gravitational potential well. Similar processes are also proposed to explain the escape of atomic hydrogen in Early Venus and Early Earth \citep{Kasting+Pollack1983IC, WatsonEt1981}. For this study, we implement irradiation-driven mass loss in MESA using the prescriptions of \citet{Murray-ClayEt2009ApJ}. Note that in this work we only focus on implementing hydrodynamic mass loss; we do not directly treat Jeans escape, direct blow-off, or ``photon-limited" escape \citep{Owen&Alvarez2015ApJ} which each may be relevant in other regimes.  

At low levels of irradiation, the mass loss is assumed to be energy limited. We follow the energy limited escape formulation first described by \citet{WatsonEt1981} then studied by \citet{LammerEt2003ApJ} \citet{ErkaevEt2007A&A}, \citet{ValenciaEt2010A&A}, \citet{LopezEt2012ApJ}, and \citet{LugerEt2015ApJ}.  %This scheme has also been studied extensively by prior work such as \citet{ValenciaEt2010A&A}, \citet{LopezEt2012ApJ}, and \citet{LugerEt2015ApJ}. 
The energy-limited mass loss rate is given by:

\begin{equation} 
\frac{{\rm d} M_p}{{\rm d} t} = - \frac{\epsilon_{\rm EUV} \pi F_{\rm EUV} R_{\rm p} R_{\rm EUV}^2}{G M_p K_{\rm tidal}}, \label{elim}
%\approx -\frac{{{\rm d} E_{\rm EUV}}/{{\rm d} t}}{{{\rm d} E_{\rm potential}}/{{\rm d} m}}
\end{equation}

\noindent where $\epsilon_{\rm EUV}$ is the mass loss efficiency (i.e. the fraction of incident EUV energy that contributes to unbinding the outer layers of the planet), which depends on atmospheric composition and the EUV flux. Here we adopt a mean efficiency value of 0.1 \citep{JacksonEt2012MNRAS,LopezEt2012ApJ} unless otherwise stated. $F_{\rm EUV}$ is the extreme ultraviolet energy flux from the host star impinging on the planet atmosphere. $R_{p}$ and $M_p$ are planet radius at optical depth $\tau_{\rm visible} = 1$ (in the visible) and the total mass of the planet respectively. $G$ is the gravitational constant. $R_{\rm Hill} \approx a \left({M_p}/{3 M_{*}}\right)^{1/3}$ represents the distances out to which the planet's gravitational influence dominates over the gravitational influence of the star. 

Note that all our models we assume $R_{\rm Hill}$ to be located well within the exobase where the particle mean free path and atmospheric scale height are comparable. $K_{\rm tidal}$ (a factor that depends on the ratio of $R_{\rm Hill}$ and $R_{\rm EUV}$) corrects for tidal forces, which modify the geometry of the potential energy well and decrease the energy deposition needed to escape the planet's gravity \citep{ErkaevEt2007A&A}.

Finally, $R_{\rm EUV}$ is distance from the center of the planet to the point where the atmosphere is optically thick to EUV photons. To calculate $R_{\rm EUV}$, which changes with time, we first approximate the difference between $\tau_{\rm visible}=1$ and $\tau_{\rm EUV}=1$ (the photo-ionization base) with

\begin{equation}
R_{\rm EUV} \approx R_p + H \ln \left(\frac{P_{\rm photo}}{P_{\rm EUV}}\right)
\end{equation}

\noindent where  $H = (k_{\rm B} T_{\rm photo})/(2 m_H g)$ is the atmospheric scale height at the photosphere (the factor of 2 in the scale height equation denotes the molecular form of hydrogen in this regime). $P_{\rm photo}$ and $T_{\rm photo}$ are the pressure and temperature at the visible photosphere. 

Following \citet{Murray-ClayEt2009ApJ}, we estimate the pressure at $\tau_{\rm EUV}=1$ from the photo-ionization of hydrogen, $\sigma_{\nu_0}=6\times 10^{-18} \left(h\nu_0/13.6~\mathrm{eV}\right)^{-3}~\mathrm{cm^2}$ as $P_{\rm EUV}\approx \left(m_HGM_p\right)/\left(\sigma_{\nu_0}R_p^2\right)$, adopting a typical EUV energy of $h\nu_0=20~\mathrm{eV}$ instead of integrating over the host star spectrum.   

At high EUV fluxes $\left(\gtrsim 10^4~\mathrm{erg\,s^{-1}\, cm^{-3}}\right)$, radiative losses from Ly$\alpha$ cooling become important, mass loss ceases to be energy limited \citep{Murray-ClayEt2009ApJ} and a constant mass loss efficiency parameter assumption no longer holds. In this regime, photo-ionizations are balanced by radiative recombinations and radiative losses maintain the temperature of the wind at $T_{\rm wind}\sim 10^4~\mathrm{K}$. This radiation-recombination limited mass loss rate is approximated by,

\begin{equation}
    \left.\frac{{\rm d} M_p}{{\rm d} t}\right|_{rr-lim} = - \pi \left(\frac{GM_p}{c_s^2}\right)^2c_sm_H\left(\frac{F_{\rm EUV}GM_p}{h\nu_0\alpha_{rec}R_{\rm EUV}^2c_s^2}\right)^{1/2}e^{\left(2-\frac{GM_p}{c_s^2R_{\rm EUV}}\right)}, \label{eqn:rr}
\end{equation}

\noindent where $c_s=\left({2k_BT_{\rm wind}}/{m_H}\right)^{1/2}$ is the isothermal sound speed of the fully ionized wind, and $\alpha_{\rm rec}$ is the radiative recombination coefficient at $10^4$~K $\left(2.7\e{-13}~\mathrm{cm^{3}\, s^{-1}}\right)$. Equation~\ref{eqn:rr} is identically Equation~20 of \citet{Murray-ClayEt2009ApJ}, but with the dependencies on planet mass and surface gravity explicitly preserved
\citep[see also][]{Kurokawa&Nakamoto2014ApJ}.

For the EUV luminosity of the planet host star (assumed sun-like), we adopt XUV evolution model of \citet{RibasEt2005}.
At each time step, we evaluate both the energy-limited or radiation-recombination limited mass loss rates and impose the lesser of the two on the MESA planet model.

\subsection{Heavy Element Interior}
\label{subsec:core}

We incorporate two main updates to MESA to more realistically model planet heavy element interiors. First, as described in Section~\ref{subsec:entropy}, we use the models of \citet{RogersEt2011ApJ} to set the core radius and bulk density for a specified heavy element interior composition and mass. 

In this work, we do not model the variations in the core radius as a function of time or pressure over-burden. This assumption is appropriate for the low-mass planets that are the focus of this paper. However, the compression of the planet heavy-element interior due to the pressure of the surrounding envelope starts to be significant for cases in which the pressure at the boundary of the core and envelope exceeds $\sim 10^{10}$~Pa \citep[e.g.,][]{MordasiniEt2012A&A}.

Second, we incorporate into MESA a time-varying core luminosity to account for the heavy-element interior's contribution to the envelope energy budget. This contribution may be negligible in hot Jupiters, but is more significant in cases where the core represents a substantial fraction of the total planet mass.
For this reason, the energy budget for planets $M_p \la 20~M_{\rm \oplus}$ is more significantly influenced by the presence of a core. 

The core luminosity, $L_{\rm core}$, (i.e., the energy input to the base of the envelope from the heavy element interior) is commonly modeled as, 

\begin{equation} 
L_{\rm core} = - c_vM_{\rm core}\frac{{\rm d} T_{\rm core}}{{\rm d} t} + L_{\rm radio}.
\label{eqn:Lcore}
\end{equation}

\noindent The first term on the right hand side of Equation~\ref{eqn:Lcore} accounts for the thermal inertia of the core. Therein, $c_v$ is the effective constant volume heat capacity of the core (in $\mathrm{ergs\,K^{-1}g^{-1}}$), $M_{\rm core}$ is the mass of the planet's core, and $dT{\rm core}/dt$ is time derivative of the effective (mass-weighted) core temperature. We take the Lagrangian time derivative of the temperature at the base of the planet envelope as an approximation to $dT{\rm core}/dt$. The most uncertain factor is $c_v$, which could vary depending on the composition of the core and the presence of spatial composition gradients or thermal boundary layers. For the ice-rock core, we adopt a value of 1.2 J K$^{-1}$ g$^{-1}$. And the rocky core 1.0 J K$^{-1}$ g$^{-1}$ \citep{GuillotEt1995ApJ}. The second term on the right hand side of Equation~\ref{eqn:Lcore}, $L_{\rm radio}$, represents the contribution of the decay of radio nuclei to the core luminosity.

\begin{displaymath}
L_{\rm radio}= \chi M_{\rm core}\sum_{i}H_{\rm initial, i}e^{-\lambda_i t}.
\end{displaymath}

\noindent Above, $\chi$ is the mass fraction of "chrondritic" material in the planet heavy element interior; this factor is 1 for the Earth-like core composition and 0.33 for the ice-rock core composition. $\lambda_i$ is the decay rate constant, and $H_{\rm initial, i}$ is the initial rate of energy released (per unit mass of rocky material) at $t=0$ by the decay of the $i$th nuclide. The important long lived radioactive nuclides are  $^{232}$Th, $^{238}$U, $^{40}$K,  and $^{235}$U and their half-lives are respectively $1.405\e{10}$, $4.468 \e{9}$, $1.26\e{9}$, and $7.04\e{8}$ years. We use the chrondritic abundances and initial energy production rates from \citet{Hartmann2004}.  % $L_{\rm radio}$ represents the energy deposited in the core by unstable atomic nuclei emitting ionizing particles and radiation.

\section{Model Benchmarking} 
\label{sec:benchmark} 

Due to the extensive nature of our adaptations to MESA, both in the default parameters and the subroutine modules, there is a need to ensure the that our calculations are consistent with those in current literature. Thus our numerical results are first presented with a series of benchmarking exercises. 

\hfill \break

\subsection{Thermal Evolution Benchmarking}
\label{sec:benchmarking}

We first simulate the thermal evolution of Jovian and sub-Neptune planets without the effects of atmospheric escape. 

We run simulations of high mass ($M_p\geq17~M_{\oplus}$) gas giant planets to compare to the models of \citet{FortneyEt2007ApJ} (Table~\ref{tab:Fort07}). %(Table 1). %Check reference to table. does this refer to our Table 1 or Fornet et al. Table 1?
These simulated planet orbit Sun-twins, have heavy element interiors consisting of 50\% rock and 50\% ice by mass, and have no heating from the core $\left(L_{\rm core}=0\right)$ \citep[as in ][]{FortneyEt2007ApJ}.

By ages of 1 Gyr, the planet radii predicted by MESA typically agree with the tabulated radii from \citet{FortneyEt2007ApJ} to within 3\% and to better than 7\% for all planet masses and orbital separations. Radius offsets exist at younger ages (100-300 Myr), as would be expected due to the differences in how we set the initial conditions.
A crucial difference between our approaches is how we model the absorption of stellar flux in the atmosphere. \citet{FortneyEt2007ApJ} use a grid of self-consistent radiative-convective equilibrium atmospheric
structure models computed following a correlated-K approach, whereas we use a semi-grey atmospheric boundary condition based on the modified $T(\tau)$ relation of \citet{Guillot2010A&A} (Section~\ref{subsec:atm}).
Nonetheless, we find encouraging agreement between MESA and the model planet radii of \citet{FortneyEt2007ApJ}.

For planet masses below $20~M_{\oplus}$, we benchmark our results against \citet{Lopez+Fortney2014ApJ} and \citet{Bodenheimer&Lissauer2014ApJ} (Table~\ref{tab:LF14}). These simulations have Earth-like rocky cores (70\% silicate and 30\% Fe) with heat capacity of 1.0 J K$^{-1}$ g$^{-1}$ and do not experience the effects of photo-evaporation. 

At planet masses near $20~M_{\oplus}$ and at low levels of incident irradiation, we find good agreement (within 5\%) between the planet radii predicted by MESA and the results of \citet{Lopez+Fortney2014ApJ}. The differences between the two sets of model radii are most extreme at low planet masses and high $f_{\rm env}$, though all simulations agree to within 20\%. The modeled MESA planet radii further show a stronger dependence on the irradiation flux, predicting low-mass planet radii that are larger than \citet{Lopez+Fortney2014ApJ} at $1000~F_\oplus$, but smaller than those of \citet{Lopez+Fortney2014ApJ} at $10~F_\oplus$. 

As for the comparisons with \citet{Bodenheimer&Lissauer2014ApJ}, we look at their Table 2 runs with the accretion cutoff times at 2 Myr (runs 2H, 1, and 0.5). At $T_{\rm eq} = 500$~K and 200~K and ages of 4 Gyr, the final planet radii computed by MESA and by \citet{Bodenheimer&Lissauer2014ApJ} typically agree within 3\% and in all cases agree to better than 10\%.

%Thermal evolution of small planets with ice-rock cores (not Benchmarking)
In addition to the Earth-composition heavy element interiors that are used for benchmarking purposes, we also simulate the same grid of planet masses, envelope mass fractions, and orbital separations with heavy element interiors that are 70\% ice and 30\% rock by mass (Table~\ref{tab:lowMpicerock}). Notice the decisive role that changing the core composition has on the overall planetary radii. As expected, keeping all other parameters identical, the simulated planets with ice-rock cores have larger radii than those with rocky cores. 
Ice-rock cores have a smaller radio-luminosity than rocky cores due to their smaller mass fractions of radioactive nuclides.
%Less heat gets dissipated by ice-rock cores because of the higher thermal heat capacity and a smaller fraction of radioactive elements in the core.
However, this factor is greatly compensated by the fact that ice-rock cores have a lower densities and surface gravities than rocky cores. We shall see that this increase of planet radii with the replacement of ice-rock heavy-element interiors have important implications for their H/He envelope survival rates.

\subsection{Mass-Loss Evolution Benchmarking}
\label{subsec:compare}

In Section~\ref{sec:benchmarking}, we modeled planet thermal evolution in the absence of mass loss and compared the MESA-simulated planet radii to \citet{FortneyEt2007ApJ} and \citet{Lopez+Fortney2014ApJ} (Tables~\ref{tab:Fort07} and \ref{tab:LF14}). Here, we turn to benchmarking our evolution calculations including the full evaporation prescription described in Section~\ref{subsec:massloss}.

%Mass-loss rate calculations
We follow \citet{Murray-ClayEt2009ApJ} in calculating the mass loss rates (for hot Jupiters). For a 0.7 $M_{\rm J}$ planet with radius 1.4 $R_{\rm J}$ located at 0.05 AU around a solar-mass (G-type) star, they computed a present mass-loss rate of ${\sim} 4\e{10}~\mathrm{g\,s^{-1}}$ and a maximum mass loss rate of ${\sim} 6\e{12}~\mathrm{g\,s^{-1}}$. Our evolutionary calculations yield a close result of $ 5.5\e{10}~\mathrm{g\,s^{-1}}$ at 4.5 Gyr, although our modeled planet radius at this time is only 1.2 $R_{\rm J}$ since we do not include any hot Jupiter inflation mechanisms in our simulations.
Additionally, these mass loss rates agree qualitatively well with more detailed calculations including magneto-hydrodynamic effects (e.g. \citealt{VidottoEt2015MNRAS} \citealt{ChadneyEt2015}, \citealt{TripathiEt2015Arxiv} ). Our mass-loss rate compared to these studies do not differ by more than an order of magnitude at old ages for planets ${\sim} 1~M_{\rm J}$.

%Full evolutionary considerations
We also compared our coupled thermal-mass loss evolution calculations against previous planet evolution studies. \citet{Owen&Wu2013ApJ} found that a 318 $M_{\oplus}$ planet with a $15~M_{\oplus}$ core at 0.025 AU lost ${\sim} 0.51\%$ of its total mass over 10 Gyr. In our simulations, a planet with the same mass, composition, and orbital separation lost 5~$M_{\oplus}$ (or ${\sim} 1.81\%$ of its total mass) over the same time span of 10 Gyr. This discrepancy may be attributed to the fact that we use a fixed mass-loss efficiency factor of $\epsilon_{\rm EUV} = 0.1$, whereas \citet{Owen&Wu2013ApJ} perform a more detailed calculation of the mass loss rate. In performing a full calculation of the hydrodynamics of X-ray driven escape, \citet{Owen&Jackson2012ApJ} concluded that Jovian mass planets drive the least efficient winds, and that the effective approximate escape efficiency for this particular planet scenario is ${\sim}0.01$.  %They explain this deviation from the nominal 0.1 value by the decrease in the dynamical time-scale for low-mass planets, while the thermal time-scale remains roughly constant throughout.%Double check
Assuming $\epsilon_{\rm EUV} = 0.01$ in our simulation, the cumulative mass lost agrees very well with the results of \citet{Owen&Jackson2012ApJ}, with our simulation losing ${\sim} 0.57\%$ of its total mass.

\citet{Lopez&Fortney2013ApJ} simulated mass loss from a $ 320~M_{\oplus}$ Jupiter-mass planet with a $64~M_{\oplus}$ core at 0.033 AU (roughly $1000~F_{\oplus}$). They found that the planet lost less than 2\% of its total mass. When we model an identical planet we find a total of 1.1\% of mass lost over 10 Gyr. The fact that we find a lower cumulative mass loss is to be expected because \citet{Lopez&Fortney2013ApJ} assume energy limited escape throughout the entire planet evolution, which would lead to an overestimation of the total mass lost.

\begin{figure}[thb] %different options for where to place figure
\begin{center}
\includegraphics[width=1.0\columnwidth]{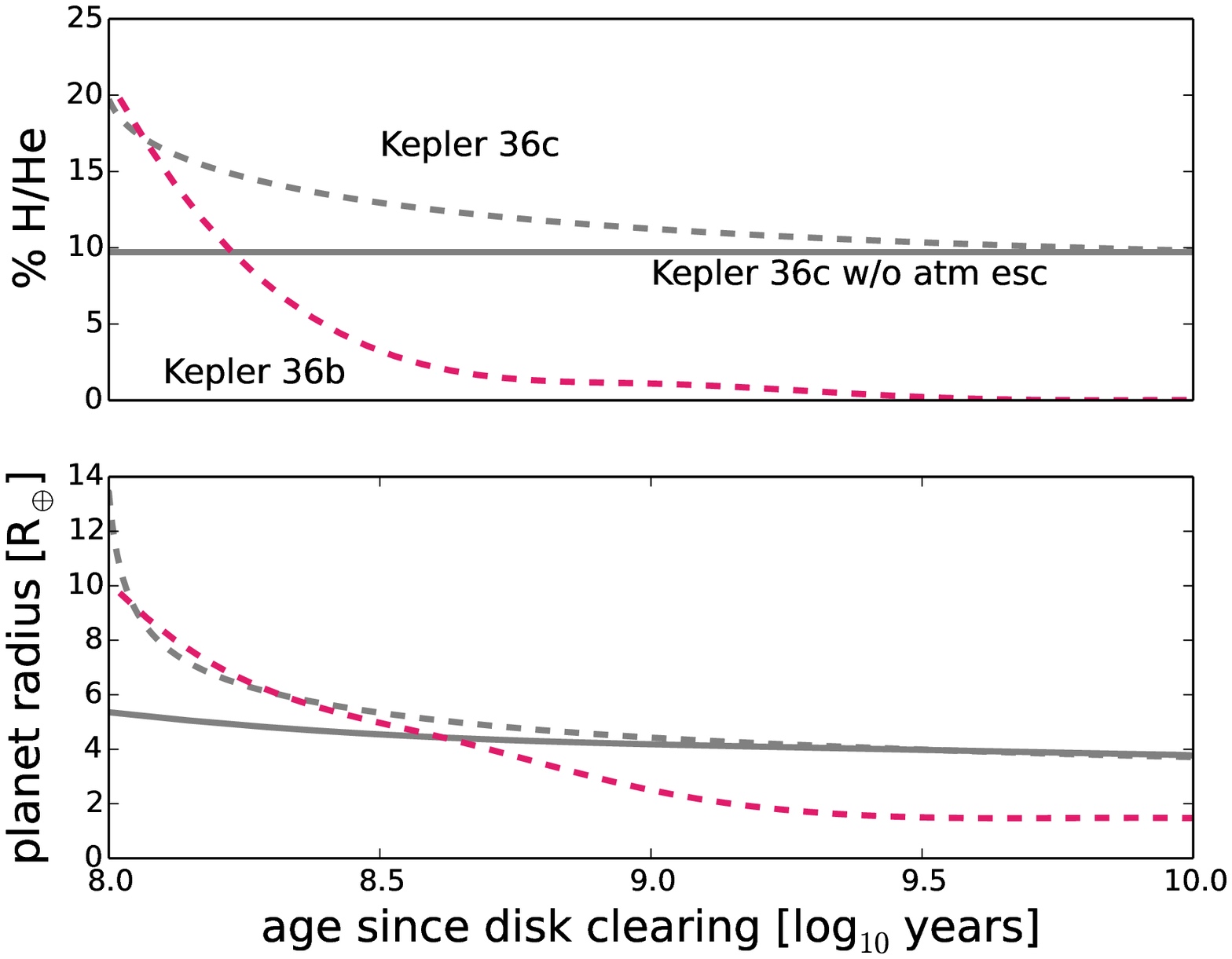}
\caption{\label{fig:kepler36}
Plot of the thermal/mass-loss evolution of Kepler-36 b and c, with radius and envelope mass fraction as a function of time. This plot is designed to be compared with Figure~1 from  \citet{Lopez&Fortney2013ApJ}. The dashed curves have initial compositions and masses (Kepler-36c: $f_{\rm env} = 22\%$, and $M_p = 9.41 M_{\oplus}$, Kepler-36b: $f_{\rm env} = 22\%$, and $M_p = 4.45 M_{\oplus}$) that \citet{Lopez&Fortney2013ApJ} used it their evolution calculations to match the current properties of the planets inferred by \citep{CarterEt2013}. The solid curve is a simulation that begins with the inferred {\it current} composition and evolved in the absence of photo-evaporation. Note the relatively small final radii difference (${\sim} 1.35\%$) between solid and dashed curves for Kepler-36c. }
\end{center}
\end{figure}

%Paragraph on lower mass planets
For simulations of mass loss from planets in the sub-Neptune/Super-Earth regime, we compare with figure 3 in \citet{Owen&Wu2013ApJ}. This is a (initially) $20~M_{\oplus}$ planet with a $12~M_{\oplus}$ core. With their X-ray and EUV driven evaporation scheme, \citet{Owen&Wu2013ApJ} found that the planet ended up with about $13~M_{\oplus}$. We find a slightly higher final mass value of $13.45~M_{\oplus}$. This difference may be partly due to the fact that \citet{Owen&Jackson2012ApJ} more carefully treats the difference in EUV and X-ray driven wind scenarios. % our analytic evaporation is primarily in the EUV regime, as opposed to the combined (X-ray and EUV) regime \citep{Owen&Jackson2012ApJ}. 
We also compute the mass loss evolution of the Kepler-36 system to reproduce Figure 1 from \citet{Lopez&Fortney2013ApJ} (Figure~\ref{fig:kepler36}). These simulations start with the initial compositions and masses  (Kepler-36c: $f_{\rm env} = 22\%$, and $M_p = 9.41 M_{\oplus}$, Kepler-36b: $f_{\rm env} = 22\%$, and $M_p = 4.45 M_{\oplus}$) that \citet{Lopez&Fortney2013ApJ} used it their evolution calculations to match the current properties of the planets inferred by \citep{CarterEt2013}. 
Our calculated evolution tracks of both planets agree qualitatively well with those from \citet{Lopez&Fortney2013ApJ} (Figure~\ref{fig:kepler36}); we also find that 36b ends with a density higher than 36c by a factor of 8 despite the two beginning with the same composition. Note however, that the predicted planet radii in our models are slightly above the measure radii (for instance, we predicted 3.75 $R_{\oplus}$ as opposed to the measured 3.67 $R_{\oplus}$ for 36c).

To summarize Sections~\ref{sec:benchmarking} and \ref{subsec:compare}, we find that our planet evolution results are in general agreement with those in published literature. We also find that the majority of the discrepancies can be attributed to differences in our evaporation schemes and in the way the initial starting conditions are specified.

\begin{figure}[h] %different options for where to place figure
\begin{center}
\includegraphics[width=1.0\columnwidth]{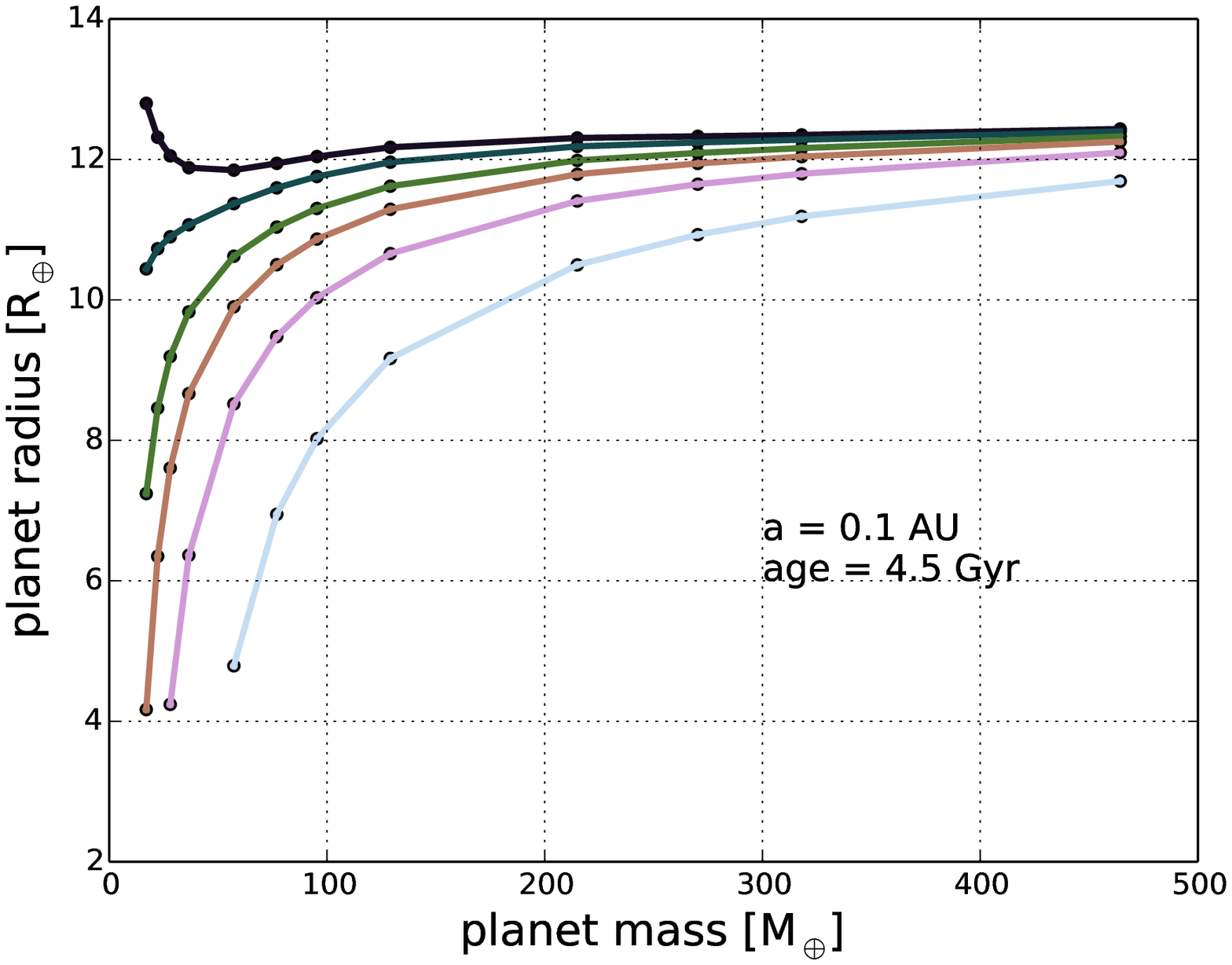}
\includegraphics[width=1.0\columnwidth]{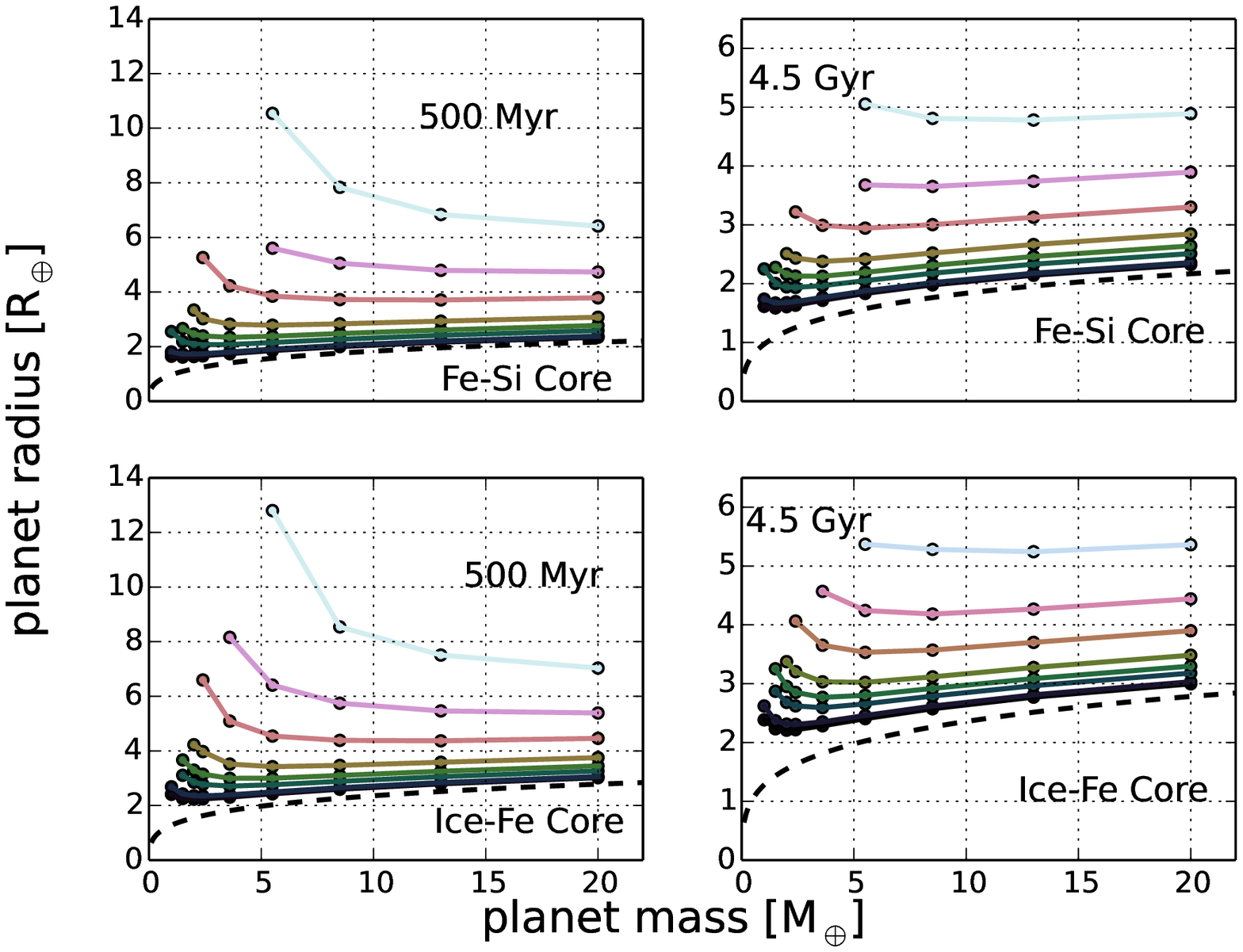}
\caption{\label{fig:rvm}
Planetary mass-radius relationship varying core mass or envelope mass fraction in the absence of mass-loss at 0.1 AU. For the top panel, the core mass values from top to bottom are 0, 5, 10, 25, 50, $100~M_\oplus$. Notice an ``ultra-low" density region where Neptune-mass planets attain sizes well above that of Saturn. For the bottom panel, compare particularly between the radii of rocky interior versus ice-rock interior simulations. The $f_{\rm env} = M_{\rm env}/M_{\rm tot}$ values from bottom to top is 0.05, 0.1, 0.5 ,1.0, 2, 5, 10, 15\%. A more exhaustive table is included in Appendix A.}
\end{center}
\end{figure}

\section{Calculations \& Results}
\label{sec:results}

We now turn to applying MESA, with the extensions discussed in Section~\ref{sec:meth}, to simulate the evolution of planets with H/He envelopes. We explore some of the important attributes of H/He-laden planets including ultra-low density configurations, the (in)dependence on evolution history of the current planetary radii and compositions, the role of evaporation in determining planet survival lifetimes as a function of composition, and a synthetic planet population generated by our models.

\subsection{Creating Mass-Radius-Composition Relations with MESA}
\label{subsec:MR}

Theoretical planetary mass-radius relationships have long been a helpful tool in aiding interpretation and characterization efforts. To produced mass-radius-composition relations (Figure~\ref{fig:rvm}), we evolve planets spanning a mass range of $2~M_{\oplus}$ to $20~M_{\oplus}$, initial envelope mass fractions of 0.05, 0.1, 0.5 ,1, 2, 5, 10, 15, 20, 25\%, and orbital separations of 0.05, 0.075, 0.1, and 1.0~AU. We evolve planets with mass loss for 1~Gyr, and 10~Gyr time spans and record both the radii and final compositions of the planets at those ages.  

With our adaptations, MESA can simulate H/He envelopes surrounding planets down to masses of 1 $M_{\oplus}$ and H/He mass fractions down to $f_{\rm env} = 1\e{-6}$, in the absence of H/He mass loss. Below $f_{\rm env} = 1\e{-6}$ the approximation of an optically thick envelope start to break down. %send 10-5 and 10-6 and then make changes
Typically, it is harder to evolve low mass ($M_p \la 15~M_{\oplus}$) planets as $f_{\rm env}$ values become greater than ${\sim} 40\%$.

With mass loss turned on in the simulations, there are certain regimes of planet parameter space in which MESA runs into issues, specifically at low planet masses, high $f_{\rm env}$, and high levels of irradiation. 
Some of the numerical difficulties can be attributed to the fact that highly-irradiated, low-mass, low-density planets are more unstable to mass loss.
For example, a $8~M_{\oplus}$ planet within 0.045 AU orbital distance might be unable to hold on to its atmosphere even for a short (${\sim} 5,000$ yr) timescale. The time steps in the evolutionary phases are generally much larger than the nominal 1,000 years. For this reason, MESA would unable to resolve the boundary conditions between the two time steps and outputs convergence error. For these regimes, instead of losing its H/He over ${\sim} 100$ Myr these simulations crash at the start. To study low mass gaseous planets in extreme equilibrium temperatures, one would need to develop a more suitable boundary condition and initialization parameters. However, these limitations largely do not impede our effort to create mass-radius isochrons in our regimes of interest.

%Power law fit to mass-radius flux relation.
From our grid of planet simulations, we derive fitting formulae that may be used to estimate the radius of a planet at specified mass, composition, irradiation flux, and age.
Though we advocate interpolating within Tables~\ref{tab:LF14} and \ref{tab:lowMpicerock} (or directly simulating the desired planet using MESA) to derive planet radii for most applications, we provide these fitting formula for situations where a quick analytic approximation may come in handy. We fit our simulated planet radii to a model in which the logarithm of the contribution of the planet's H/He layer to the planet radius $\left(R_p-R_{\rm core}\right)$ varies quadratically with the logarithms of $M_p$, $f_{\rm env}$, $F_p$, and age. 
\begin{eqnarray}
   \log_{10}\left(\frac{R_p-R_{\rm core}}{R_{\oplus}}\right) &=& c_0 + \sum_{i=1}^{4} c_ix_i +\sum_{i=1}^{4}\sum_{j\geq i}^{4} c_{ij}x_ix_j\label{eq:Rpfit}\\
   x_1&=&\log_{10}\left(\frac{M_p}{M_{\oplus}}\right)\nonumber\\
   x_2&=&\log_{10}\left(\frac{f_{\rm env}}{0.05}\right)\nonumber\\
   x_3&=&\log_{10}\left(\frac{F_p}{F_{\oplus}}\right)\nonumber\\
   x_4&=&\log_{10}\left(\frac{t}{5~\rm{Gyr}}\right)\nonumber
\end{eqnarray}
\noindent We provide the best fitting coefficients in Table~\ref{tab:plaws}, both for planets with rocky-composition cores and planets with ice-rock cores. These expressions are valid for $10^{-4}\leq f_{\rm env} \leq 0.2$, $1\leq M_p/M_{\oplus} \leq 20$, $4\leq F_p/F_{\oplus} \leq 400$, and $100~\rm{Myr}\leq t \leq 10~\rm{Gyr}$ (and should not be extrapolated beyond these ranges). Though we initially endeavored to apply a simpler linear power-law model \citep[as in ][]{Lopez+Fortney2014ApJ}, we found non-negligible curvature in the constant envelope radius hyperplanes in $\log_{10}{M_p}$, $\log_{10}f_{\rm env}$, $\log_{10}F_p$, and $\log_{10}t$ space, motivating the addition of quadratic terms to the model. Ultimately, these quadratic fits to the rocky core and ice-rock core simulations have adjusted $R^2$ of 0.993 and 0.987 (compared to 0.914 and 0.816 for the pure powerlaw) and have root mean squared residuals in the envelope radii of 0.0347 dex and 0.0465 dex (compared to 0.122 dex and 0.133 dex for the pure powerlaw).

In the above equations, the radius of the planet H/He envelope has a steep dependence on change in envelope {\it fraction} and a much shallower dependence on age. This is because non-opacity-enhanced models cool at much higher rates during the first 100 Myr of evolution.

\begin{figure}[h] %different options for where to place figure
\begin{center}
\includegraphics[width=1.0\columnwidth]{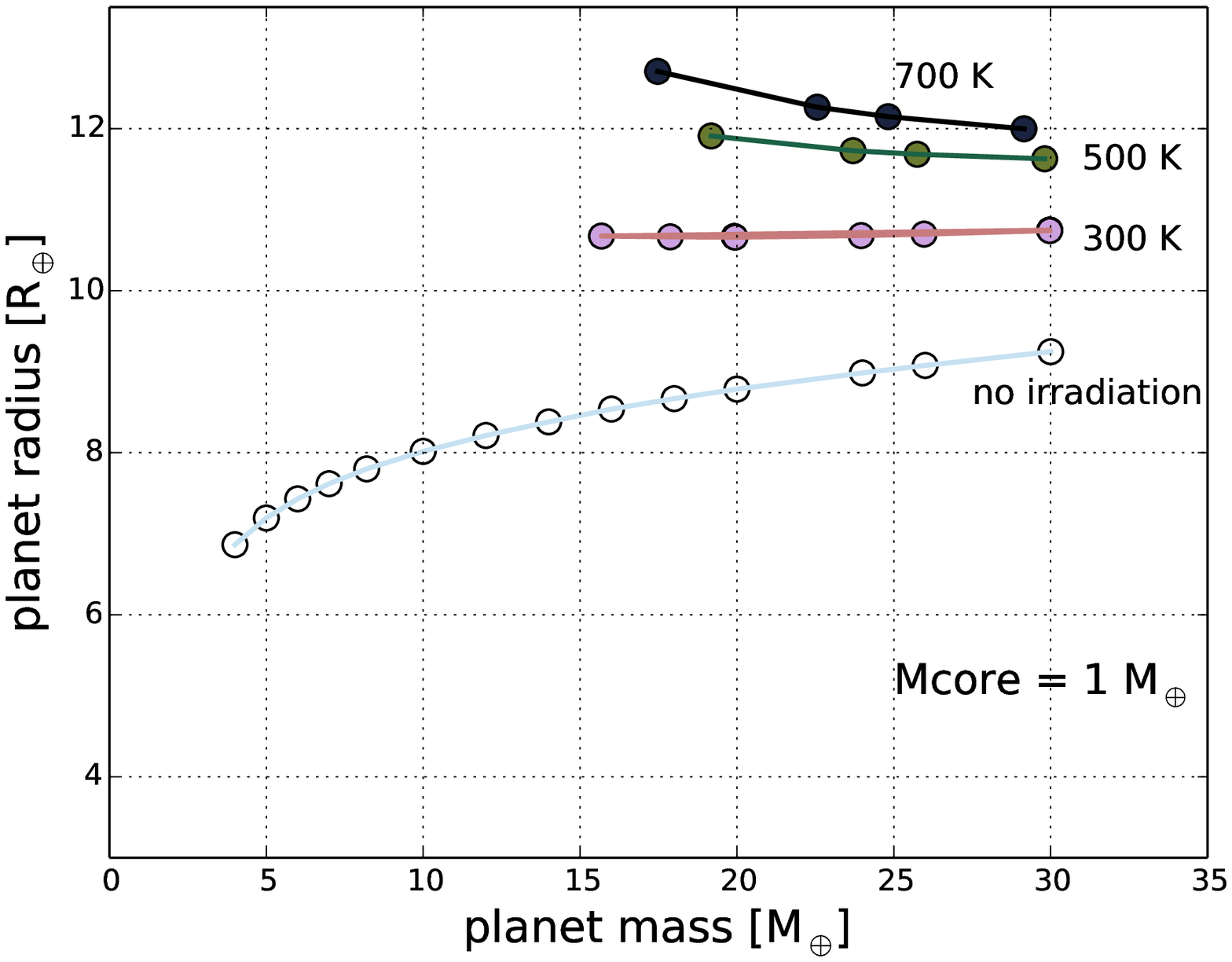}
\includegraphics[width=1.0\columnwidth]{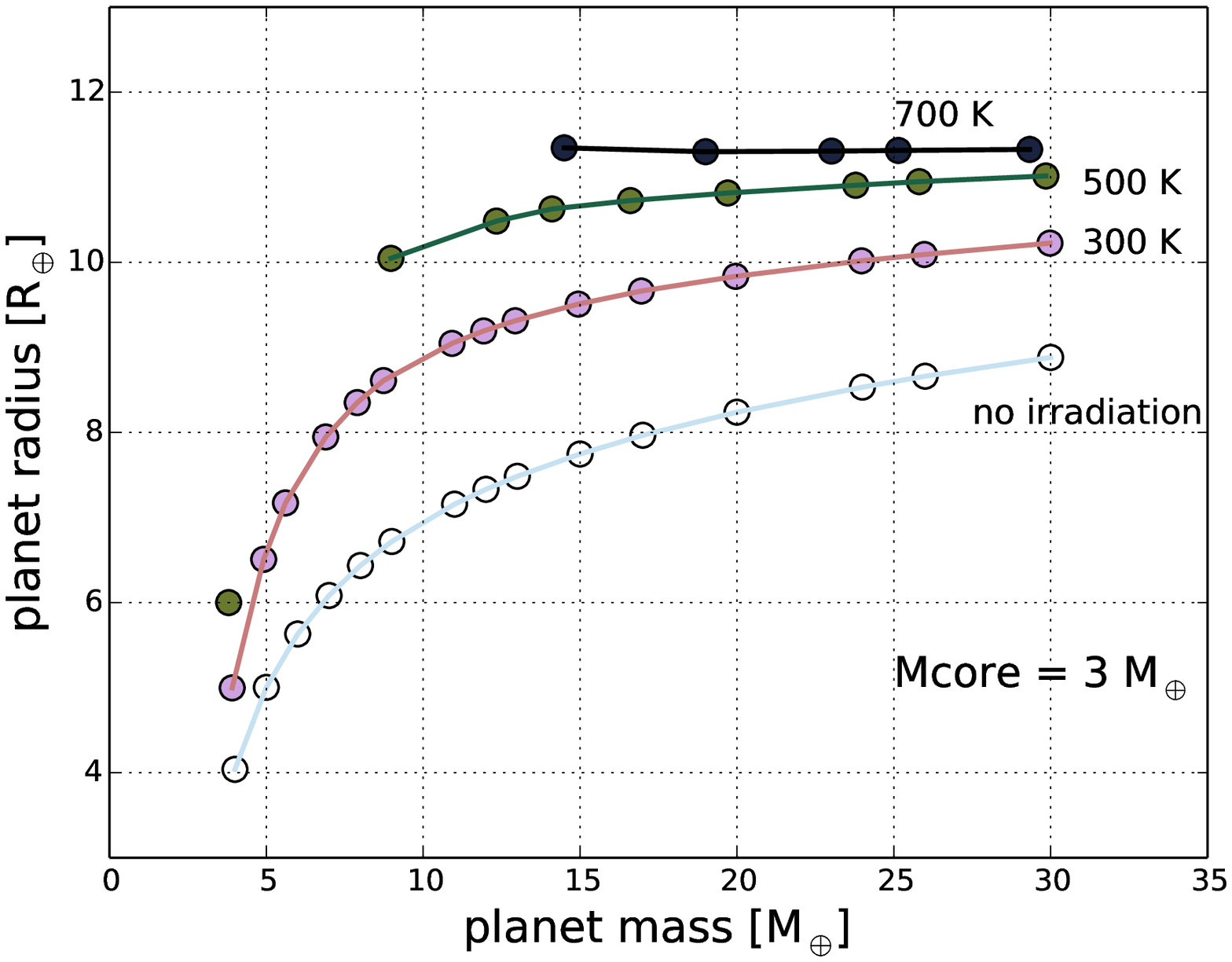}
\includegraphics[width=1.0\columnwidth]{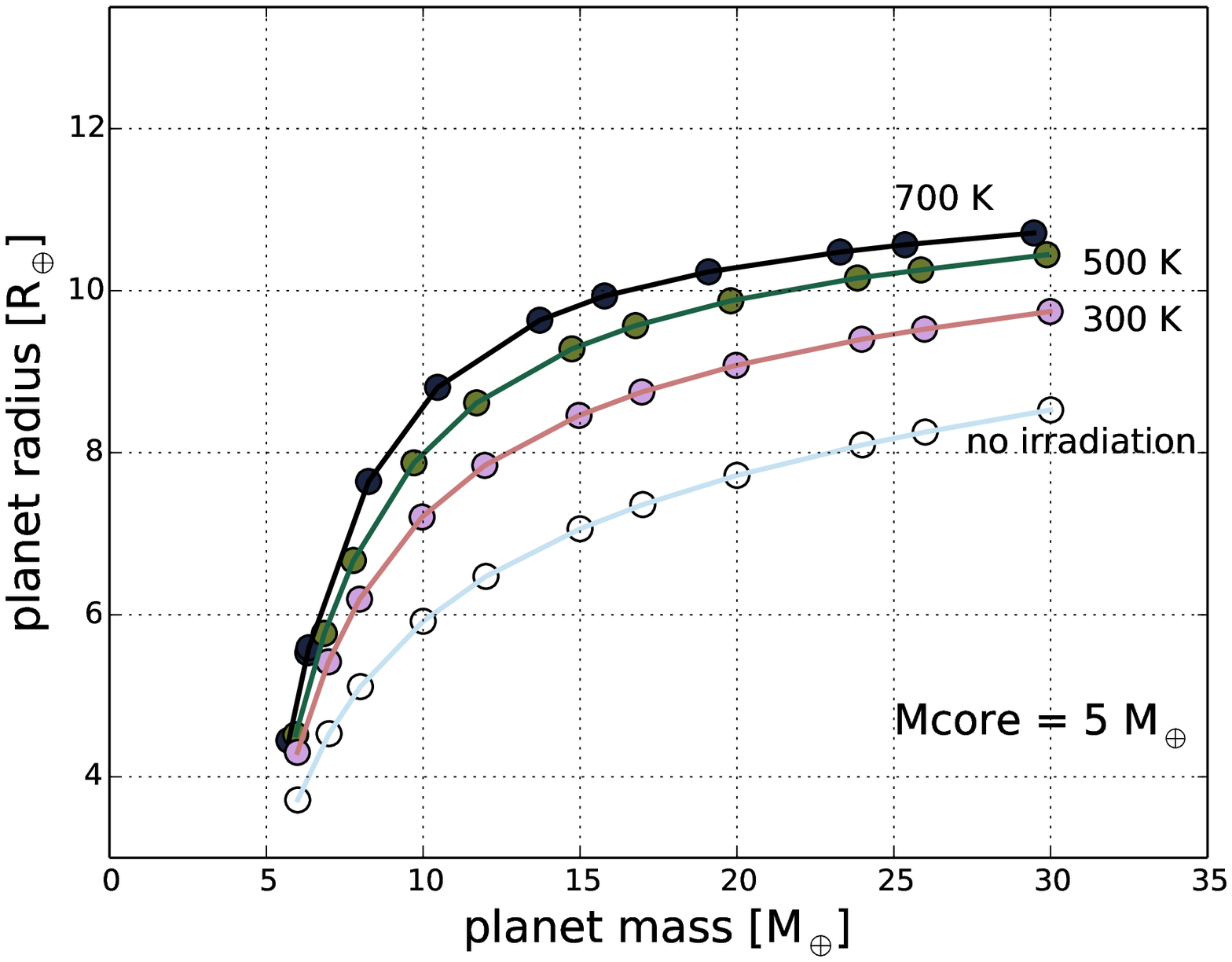}
\caption{\label{fig:low_density}
Plot of planet radii versus mass at 5 Gyr, for planets suffering evaporation. The top panel is for planets with $1~M_{\oplus}$ cores, the middle panel is for $3~M_{\oplus}$ cores, and the bottom panel is for $5~M_{\oplus}$ cores. Note that in contrast to the M-R relations from the previous sections, these are not for constant composition, but for constant core mass. This is designed to be compared with Figure~3 from  \citet{Batygin&Stevenson2013ApJ}. }
\end{center}
\end{figure}

\subsection{The Case of Ultra Low-Density Planets} 

%Motivating questions
As shown in Figure~\ref{fig:rvm}, there exist situations in which simulated sub-Saturn mass planets have larger sizes than Jupiter. These ultra low-density ``puff-ball" planets ($\rho\sim0.2 {\rm g/cm^{3}}$) are an interesting new class that challenges planet formation theories. 
\citet{RogersEt2011ApJ} showed that moderately irradiated planets low mass-planets ($3-8~M_{\oplus}$) with extended H/He envelopes can plausibly have transit radii comparable to Jupiter.  
Several measurements by {\it Kepler} provide evidence for the existence of these extremely low density planets, for example, Kepler-30d, Kepler-79 and Kepler-51 systems \citep{SanchisEt2012NAT,Jontof-HutterEt2014ApJ,Masuda2014ApJ}.

\citet{Batygin&Stevenson2013ApJ} considered these planets over a wide range of parameter space and calculated their mass-radius isochrons. However, their models assumed a constant core density ($\rho_{\rm core}=5~\rm{g/cm^3}$) independent of core mass, a luminosity independent of time, and relied on mass-loss timescale arguments to assess the survival of the planet envelope. Self-consistent calculations including a refined thermal-physical model for the planet core and coupled thermal-evaporative evolution are needed, and are now possible with the updates that we have made to MESA. Using these updates, we reproduce the mass-radius figures from \citet{Batygin&Stevenson2013ApJ}.

%Paragraph 1: 
With our coupled thermal-mass loss simulations (shown in Figure~\ref{fig:low_density}), we still find ultra low-density planets (with radii above $10~R_{\oplus}$ and masses below $30~M_{\oplus}$), increasing the confidence in the survivability of these planet configurations. With the exception of the simulations with $1~M_{\oplus}$ cores, our simulated planetary radii show good agreement with those from \citet{Batygin&Stevenson2013ApJ}.

The main differences between our mass-radius isochrons and those from \citet{Batygin&Stevenson2013ApJ} are threefold. First, our predicted planet radii are generally {\it lower} than those in \citet{Batygin&Stevenson2013ApJ}. This can be attributed to the different ways in which we set our initial starting conditions. Second, we find that planets over a slightly larger range of planet mass-core mass-incident flux parameter space are susceptible to loosing their entire envelope, with the difference being most pronounced for planets with low-mass cores ($\sim 1~M_{\oplus}$). In particular, while a $13~M_{\oplus}$ planet with a $1~M_{\oplus}$ core is expected to survive at 500~K based on the \citet{Batygin&Stevenson2013ApJ} timescale criterion, such a planet is unstable to mass loss in our simulations. Third, among the planets that manage to retain their envelopes, we find fewer instances where planet radius increases with decreasing planet mass (at constant core mass). We do still find a negative radius versus mass slope in our $T_{\rm eq} = 700 -300 $ K simulations with $M_{\rm core} = 1~M_{\oplus}$, however, this rise is radius toward smaller planet masses is less steep than that in \citet{Batygin&Stevenson2013ApJ}.

Planets with $dR_p/dM_p<0$ at constant core mass can be susceptible to runaway mass-loss \citep[e.g.,][]{BaraffeEt2008A&A}. In these configurations, removal of mass from the planet leads to an expansion of the planet radius, and further increase in the mass loss rate. This scenario is not treated in \citet{Batygin&Stevenson2013ApJ}'s instantaneous mass-loss timescale criterion. This accounts for why our simulations find a larger area of parameter space excluded by envelope mass loss in the mass-radius diagram (Figure~\ref{fig:low_density}).

\subsection{Assessing Radius as a Proxy for Composition}
\label{subsec:radiusproxy}

\citet{Lopez+Fortney2014ApJ} recasted the mass-radius relations of low mass planets as more convenient radius-\emph{H/He mass fraction} relations. However, they were focused on models that did not incorporate mass-loss in their evolutionary calculations. Such neglect is appropriate if the primary concern is to assess the effect of stellar irradiation on planetary evolution tracks. With the inclusion of hydrodynamic escape, we reassess some of their main conclusions. 

%Preamble
Figure~\ref{fig:rvm} shows that the radius of a planet with a given H/He mass fraction is quite independent of planet mass. This is because at old ages ($\ga 500$ Myr), the upturn in the radii of low mass planets at early times is largely offset by the higher cooling rate.
Such striking behavior was previously noted by \citet{Lopez+Fortney2014ApJ}, who found that the scatter in the \% H/He of their planet models at specifed radius was very low (${\sim} 0.3$ dex). This behavior suggests that planet sizes could act as direct proxy for the bulk H/He content. However, the inclusion of mass-loss may complicate matters because the planet composition does not stay constant but varies over time. %In these cases, planet radius could be a proxy for the current composition, the initial composition, or some intermediate composition.

\begin{figure}[h] %different options for where to place figure
\begin{center}
\includegraphics[width=1.0\columnwidth]{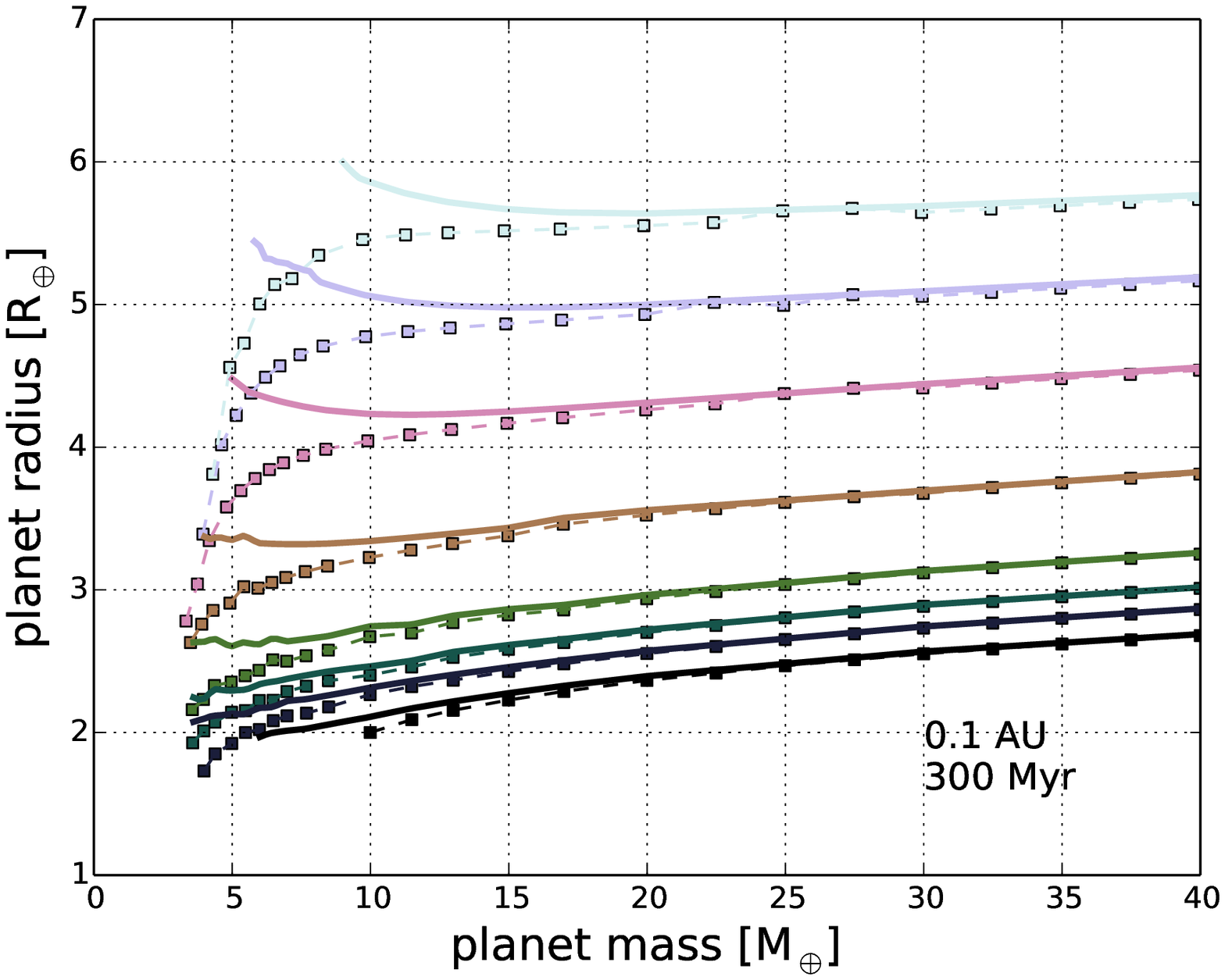}
\includegraphics[width=1.0\columnwidth]{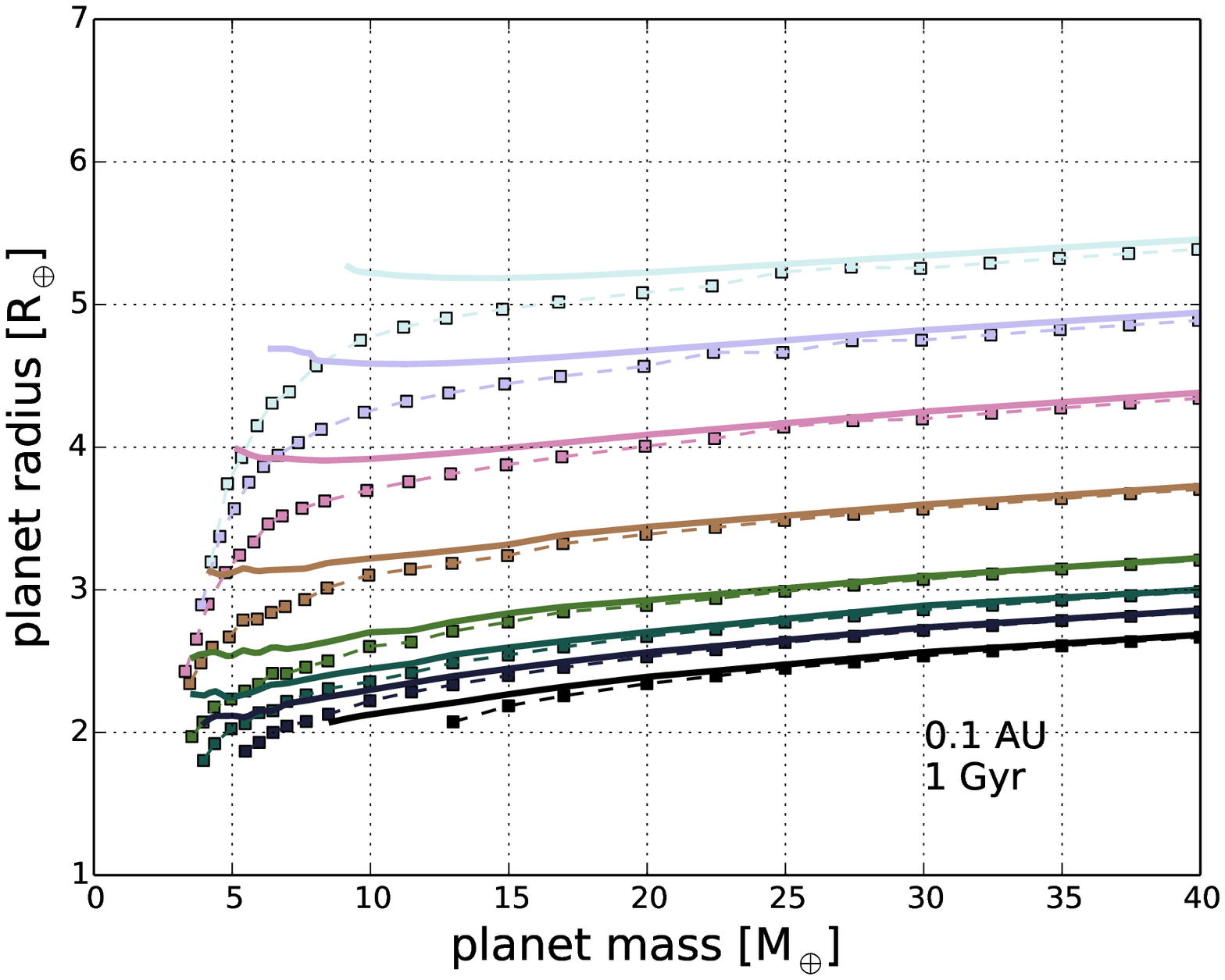}
\caption{\label{fig:RvM_proxy}
Planetary mass-radius relationships showing contours of constant initial (square markers) and final (solid curves) compositions for models experiencing evaporation. The top panel presents planets that have evolved for 300~Myr, while the bottom panel shows planets at 1~Gyr. All simulations are presented at an orbital separation of 0.1~AU. The composition sequence from bottom to top is as follows: . Note the decisive ``flattening" of the radii-composition curves for constant final composition. }
\end{center}
\end{figure}

%Our results
With the inclusion of mass-loss in our evolution calculations, we find that radius is a proxy for the \emph{current} planet composition (H/He envelope mass fraction). %("Moreover, we find that the present-day radius of a planet is a very good proxy for its initial envelope mass fraction" \citep{Howe&Burrows2015}.)\\
%This is consistent with the general lack of hysteresis in the thermal evolution of of H/He envelopes that we demonstrated in the previous section (Figure~\ref{fig:RvM_samef}).
%Figure results

Figure~\ref{fig:RvM_proxy} shows mass radius contours both at constant initial composition and at constant final compositions for planets at 0.1~AU that evolved for 300 Myr and 1 Gyr.  Despite the addition of mass-loss, the contour lines of constant final composition are flat, with radius largely independent of planet mass ($M_p \la 30~M_{\oplus}$) for planets older than $\sim800$~Myr. For example, in the mass range of 5 to $20~M_{\oplus}$, the radii of our simulated planets with 1\% H/He vary by no more than 0.5~$R_{\oplus}$, while the radii of planets having 15\% H/He vary by no more than 0.1~$R_{\oplus}$ over the same mass range. 
Interestingly, these results hold even at a closer orbital separations; at 0.05 AU; simulations that survive over 1 Gyr (typically $\ga 8~M_{\oplus}$) still have flat mass-radius curves.
For young planets $\la500$~Myr there is an upturn in the mass-radius relations (at constant final compositions) at low-masses. Radiative cooling over time decreases these ``inflation" of low mass planet sizes to which planets from 1 to $20~M_{\oplus}$ have comparable radii.

\begin{figure*}[t]
\centering
  \begin{tabular}{@{}cccc@{}}
\includegraphics[width=1.0\columnwidth]{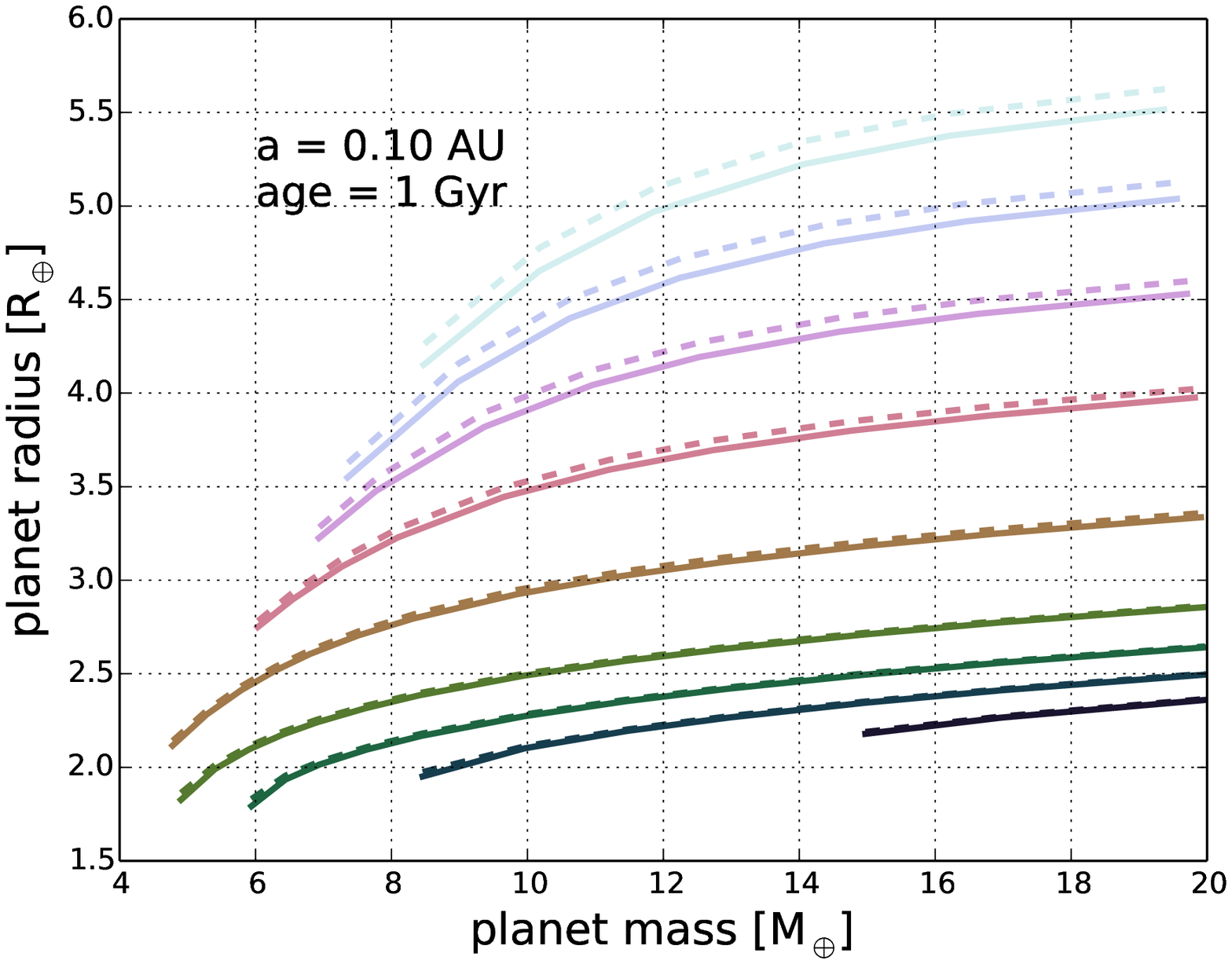} &
    \includegraphics[width=1.0\columnwidth]{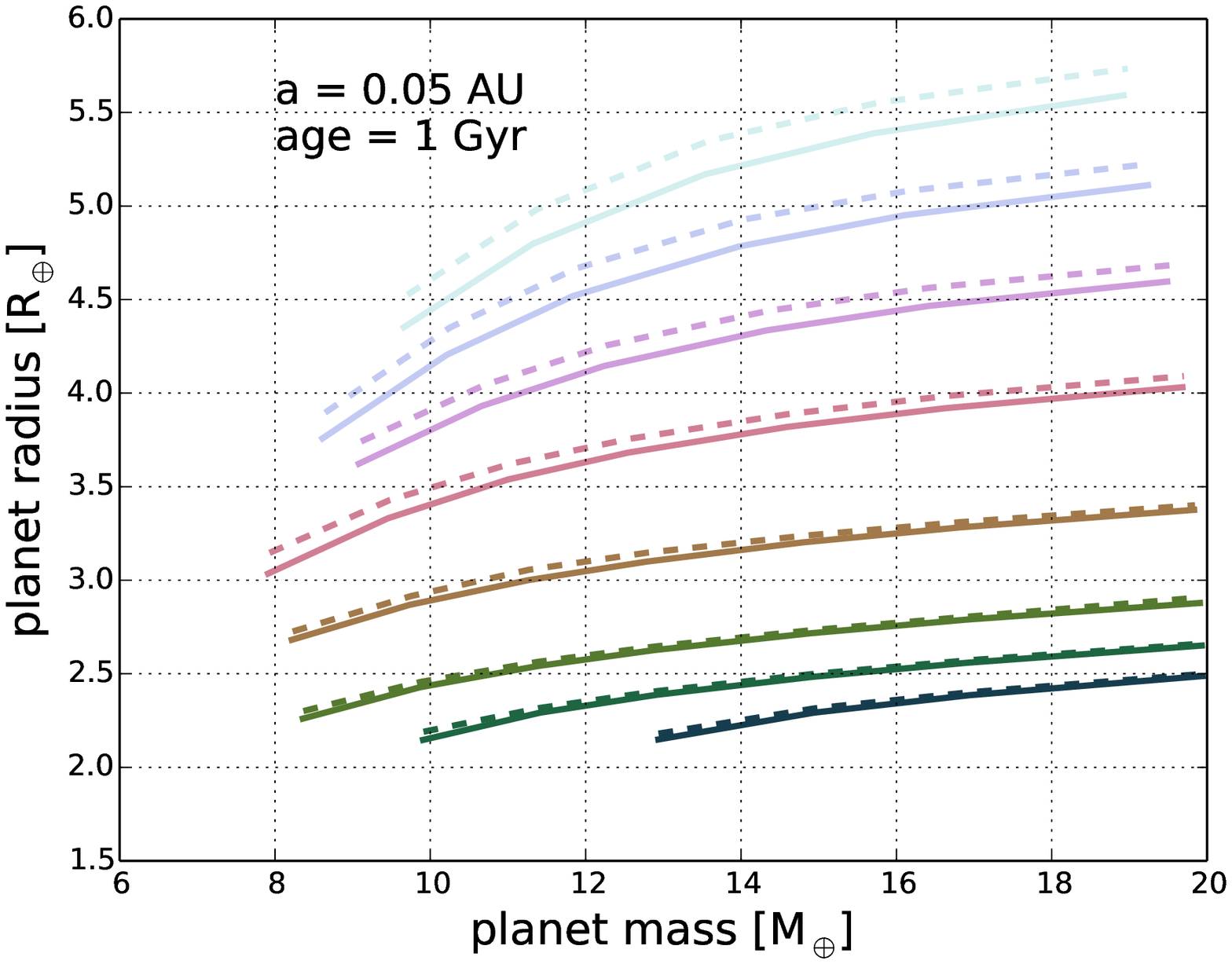} \\
\includegraphics[width=1.0\columnwidth]{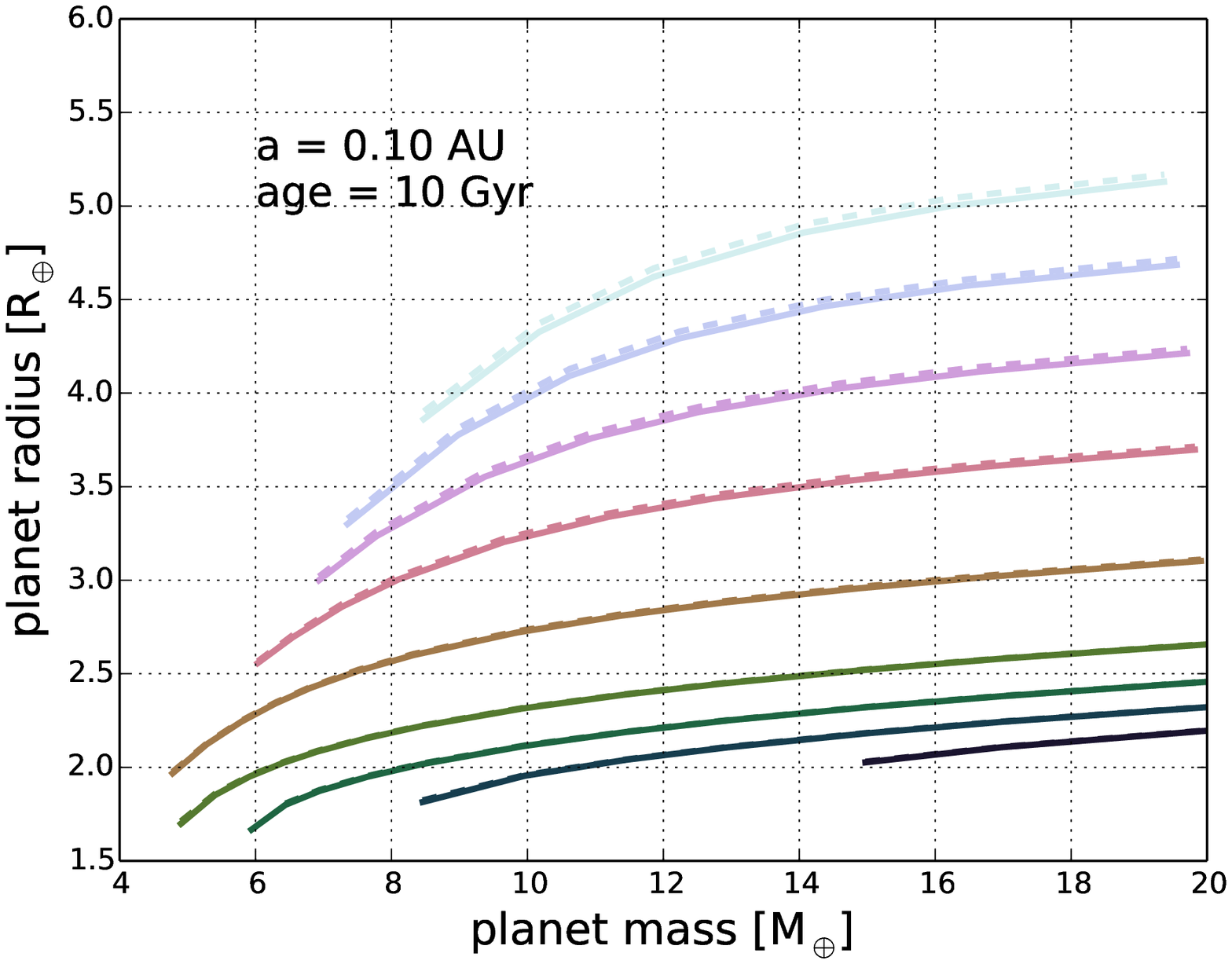}  &
    \includegraphics[width=1.0\columnwidth]{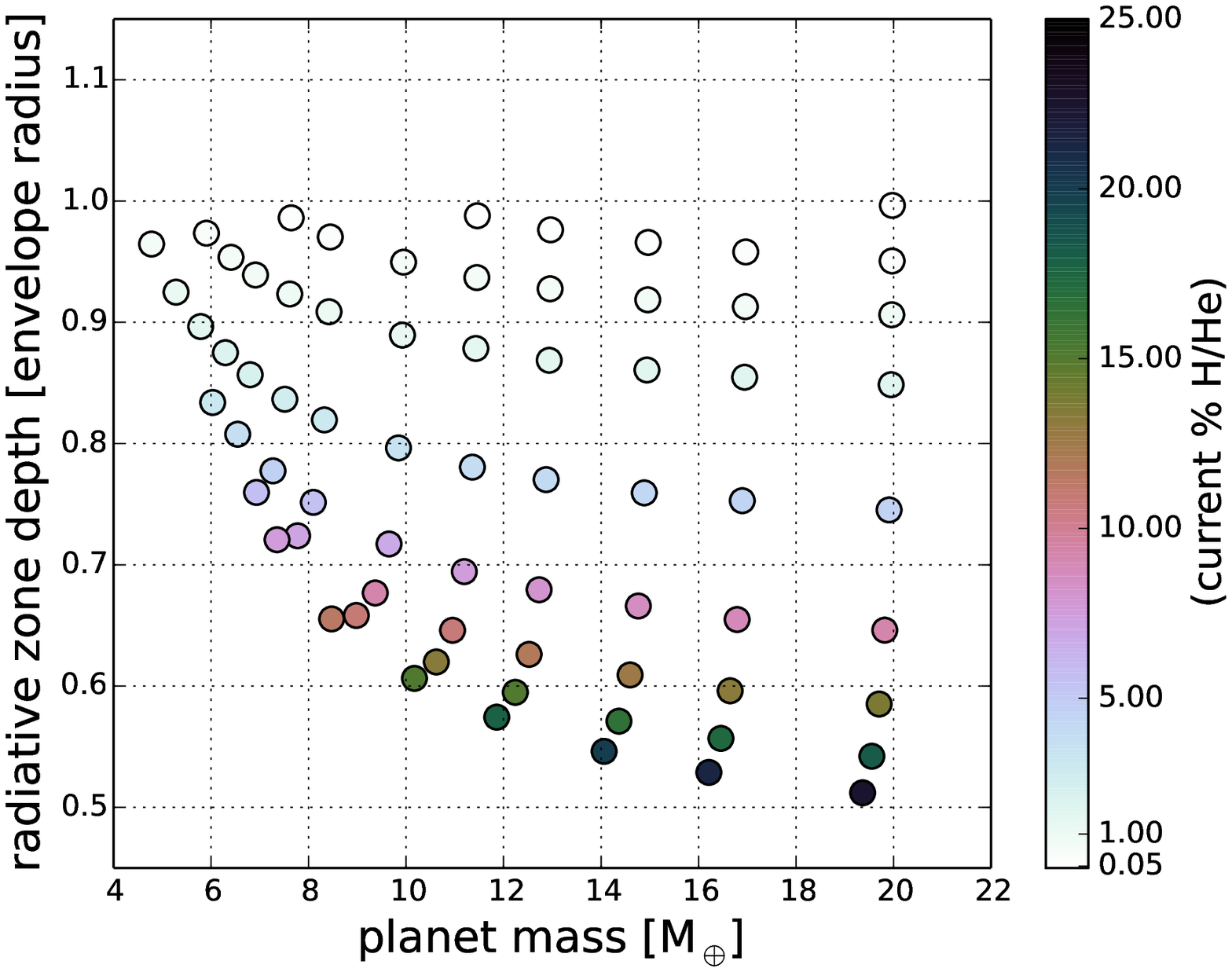} \\
  \end{tabular}
  \caption{\label{fig:RvM_samef} Quantifying the degree of hysteresis with planet evolution tracks experiencing mass-loss. The top panels show plot of planet mass-radius relations with varying initial envelope mass fractions, at an age of 1 Gyr, and at orbital distances of 0.10 AU (top left) and 0.05~AU (top right). Dotted lines represent simulations of planets experiencing mass-loss, while the solid curves represent planets (of identical instantaneous mass and composition) that evolved at constant mass. The initial envelope mass fraction values, from top to bottom, are: 25\%, 20\%, 15\%, 10\%, 5.0\%, 2.0\%, 1.0\%, 0.05\%, 0.01\%. Accompanying higher values of \% H/He in the planet composition is the increase of the discrepancy between the dotted and solid lines. This discrepancy is weakly dependent of planet mass, with lower mass simulations ($M_p \la 10~M_{\oplus}$) having more significant discrepancy (up to 3\% in planet radius). Conversely, in the regime of low H/He mass fractions, there is little difference between the radii of planets following the two different evolution pathways. The bottom left panel shows the same relations at 0.1~AU but for models with the same instantaneous H/He fractions at 10~Gyr. The bottom right panel present the fractional radiative zone depth (in units of the total radial thickness of the planet H/He envelope, $R_{\rm env}$) as function of planet mass, for 1~Gyr old planets at 0.1 AU. Notice how the outer radiative zone represents a larger fraction of the envelope radius at lower H/He mass fractions and smaller planet masses, explaining the relative small display of hysteresis in this region of parameter space.}
\end{figure*}

\subsection{(In)Dependence on Evolution History}
\label{subsec:history}

Planet evolution calculations, in principle, provide a useful mapping from planet mass, composition, age, and irradiation to planet radius. Often, mass-radius isochrones calculated neglecting mass loss from H/He envelopes \citep{FortneyEt2007ApJ,Lopez+Fortney2014ApJ,HoweEt2014ApJ}, are applied even in scenarios in which planetary evaporation processes may be significant. \citet{LopezEt2012ApJ} previously noted that evaporation only strongly affects thermal evolution in cases of extreme mass loss when the evaporation timescale becomes comparable to the thermal cooling timescale. However, it is still crucial to more quantitatively calculate the error introduced by not accounting for the full mass-loss history of a planet. In this section we explore the regimes of parameter space in which planet evolution may exhibit hysteresis, or in other words, in which the radii of planets with identical compositions, masses, and ages depend on their earlier evolution history. We first use the specific case of Kepler-36c to provide an illustrative example of moderate hysteresis. 

Figure~\ref{fig:kepler36} shows two distinct evolution tracks for Kepler-36c, which both end up at 10 Gyr, with identical final compositions and masses ($f_{\rm env} = 8.2\%$, $M_p = 8.01~M_{\oplus}$). The first evolution track (dashed line) includes atmospheric mass loss over the cumulative history of the planet, and began with an initial composition of 22\% H/He (9.41 $M_{\oplus}$ total mass). In contrast, the second evolution track (solid-line) does not include mass loss, staying at constant mass and composition throughout its lifetime. The final planet radii of the two simulations at 10 Gyr show a ${\sim} 1.35\%$ difference (3.65 vs 3.51 $R_{\oplus}$), with the non-evaporating model having a smaller radius. While this difference in the model radii appears small, it approaches the measurement uncertainty on Kepler-36c's radius ($\lesssim 2\%$). This example motivates further investigation to map out the scenarios in which planet radii are even more strongly dependent on the planet evolution history.

We simulate a large grid of planet evolution models to extend the hysteresis experiment to a wider range of planet masses, envelope fractions, and orbital separations. We use the same masses as Section~\ref{subsec:MR}, initial $f_{\rm env}$ = 25, 20, 15, 10, 5.0, 2.0, 1.0, 0.05, 0.01\%, simulating a grid of evolution calculations with mass loss turned on. We then performed a second suite of planet evolution simulations with mass loss turned off. These simulations have compositions identical to those of the mass-losing simulations at 1 and 10 Gyr. The results for 1~Gyr old planets at 0.10~AU and 0.05~AU are shown in Figure~\ref{fig:RvM_samef}. In contrast to Section~\ref{subsec:radiusproxy}, where we presented simulations with the same initial compositions, here we compare suites of simulations that begin with different initial compositions but end up with identical final compositions. 

%%%%%
With the inclusion of mass-loss, planets tend to be larger at a specified mass and instantaneous composition.
For example, at 0.05~AU and 1~Gyr ages, the radius of a planet with a specified instantaneous mass and composition can vary by up to 2.5\% (at $15~M_{\oplus}$, and $f_{\rm env}=25\%$). At ${\sim}0.1$ AU, where lower mass planets $\left(M_p \sim 5~M_{\oplus}\right)$ may retain their atmospheres (1-10\% H/He), we find overall planet radius differences of ${\sim} 1\%$. 
At a given composition, higher mass-loss rates lead to greater differences in the final sizes. As a corollary, younger simulations receiving higher irradiation with higher envelope mass fractions are marked by a greater difference in planet radii.
At older ages beyond 10~Gyr, hysteresis is almost non-existence for planets with $f_{\rm env} \la 15\%$.

%In what regime is the radius difference important/negligible: 
At lower initial $f_{\rm env}$ below 10\%, the difference between evolution computations including and disregarding mass-loss is never more than 0.5\% regardless of flux received or planet mass. At further out orbital semi-major axes of 0.4 AU, there is minimal discrepancy ($\la 0.01\%$) between simulations including and excluding mass-loss in their evolutionary computations. Nonetheless, with the exception of strongly irradiated orbital separations below ${\sim} 0.06$ AU, the difference is always within the 1\% mark. This suggests that, in most cases the error introduced by using mass-radius isochrons calculated for planets neglecting mass loss is negligible. Nonetheless, the mass loss history of a planet can introduce a systematic shift in planet mass-radius relations that can be comparable to the observations radius uncertainties for planets with $f_{\rm env}\gtrsim 10\%$ at 1 Gyr. This difference typically decreases to between 0.2 and 0.5\% at even older ages of 10 Gyr. The degree of hysteresis is only weakly dependent on planet mass.

%explain?
To get insight for the reason behind these trends, we can examine the structure of the planet H/He envelope. Since more massive planets start with higher interior entropy, removing mass from a planet's envelope leads to an interior structure that is effectively ``younger" (higher entropy) than a planet that evolved at a constant lower mass. This explains the fact that the simulated planets that experienced mass loss have systematically higher radii (at specified mass, age and composition) than the simulated planets that evolved at constant mass. Planets for which the convective H/He envelope contributes a significant fraction of the planet radius, should be more susceptible to hysteresis than planets with deep radiative envelopes. The temperature of the outer radiative zone in a planet's H/He envelope is set by the radiation incident on the planet from its host star (independent of the cooling history of the planet). It is the location of the radiative-convective boundary and the entropy of the convective zone that may depend on the planet's earlier evolution. 

The radiative zone depth in our simulated planets (at specified age and orbital separation) depends only weakly on mass but more heavily on $f_{\rm env}$ (Figure~\ref{fig:RvM_samef}). The fraction of the envelope radius that is radiative decreases with increasing H/He mass fraction. 
Planets with low H/He mass fractions (below ${\sim} 1\%$) have envelopes that are almost entirely radiative ($\ga 80 $ to $100\%$ of the envelope radius lying in the radiative zone). Planets with very low total mass ($M_p \la 7~M_{\oplus}$) also have envelopes that are almost entirely radiative.
At the other extreme, comparison of the top and bottom panels in Figure~\ref{fig:RvM_samef} indicates that in regimes where the simulated planets do exhibit hysteresis (at $f_{\rm env}\ga 10\%$, where the eventual planet radii differ by ${\sim} 0.5-1.5\%$) planets generally have more substantial convective regions, accounting for at least 30\% of the total radial extent of the planet envelope.

It is possible however, that the cause of the apparent hysteresis in our planet evolution simulations is numerical in nature. This is because our modified $T(\tau)$ atmospheric boundary condition does not conserve energy globally to machine precision. Since mass-loss has some $pdV$ work associated with it in the upper atmosphere, a fixed temperature boundary condition does not treat this correctly, and in some cases will lead to more energy input into the planet's atmosphere. In reality however, the advection due to evaporation carries some extra thermal energy {\it out} of the planet, thereby enhancing cooling. For the ``normal" EUV evaporation, the expectation is that this is a small effect, hence using a fixed $T(\tau)$ relation would not introduce significant errors. However, once the escape enters the ``boil-off" regime (e.g. \citealt{Owen&Wu2015ApJ}), then our current boundary condition would not be suitable.

Regardless of the root cause, (numerical or physical in nature), our main conclusion is unchanged. The error introduced by using mass-composition-radius isochrons calculated for planets neglecting mass loss is typically small ($\lesssim 1\%$ in planet radius).

\subsection{A Favored \% H/He mass fraction?} %Can Evolution Produce a Favored \% H/He mass fraction?
\label{subsec:tau_env}

%Intro to the observational signature
Based on the analysis of the radius distribution of a subset of {\it Kepler} planet candidates, \citet{Wolfgang+Lopez2015ApJ} found a typical H/He mass fraction of 0.7\% (with a standard deviation of ${\sim} 0.6$ dex). Ultimately the compositions of planets observed today are consequences of both the initial formation and subsequent evolution. What role might evaporative planet mass loss play in producing this 0.7\% typical H/He envelope mass fraction? %question?
%By more quantitative means, it is possible to find out whether this 0.7\% typical H/He envelope mass is another signature of evaporative mass loss.

%Introduction to simulations based on M-R diagrams
Our simulations of the coupled thermal and mass-loss evolution of low-mass planets show evidence for a non-monotonic relation between planet composition and envelope survival rate. Similar behavior has also been noted in previous planet evaporation parameter studies \citep{Lopez&Fortney2013ApJ,JinEt2014ApJ}. In Figure~\ref{fig:RvM_proxy}, we see that, by ages of 1 Gyr, low mass planets ($\lesssim6~M_{\oplus}$) with high (${\sim} 20\%$) or very low (${\sim} 0.05-0.1\%$) envelope mass fractions tend to have lower survival rates. Envelope mass fraction values in the "intermediate" range seem to have a greater probability of survival. This is seen in Figure~\ref{fig:RvM_proxy} by the fact that the mass-radius curves for $f_{\rm env }= 0.05-2\%$ extend to lower masses than the mass radius curves of both higher and lower envelope fractions.  % Specifically, \citet{Lopez&Fortney2013ApJ} attributed the change in behavior at 1\% to the altering the feedback between envelope fraction and a planets volume and therefore its mass loss rate.} 
This could provide a mechanism to imprint a preferred envelope mass on the planet population.

\begin{figure}[t] %different options for where to place figure
\begin{center}
\includegraphics[width=1.0\columnwidth]{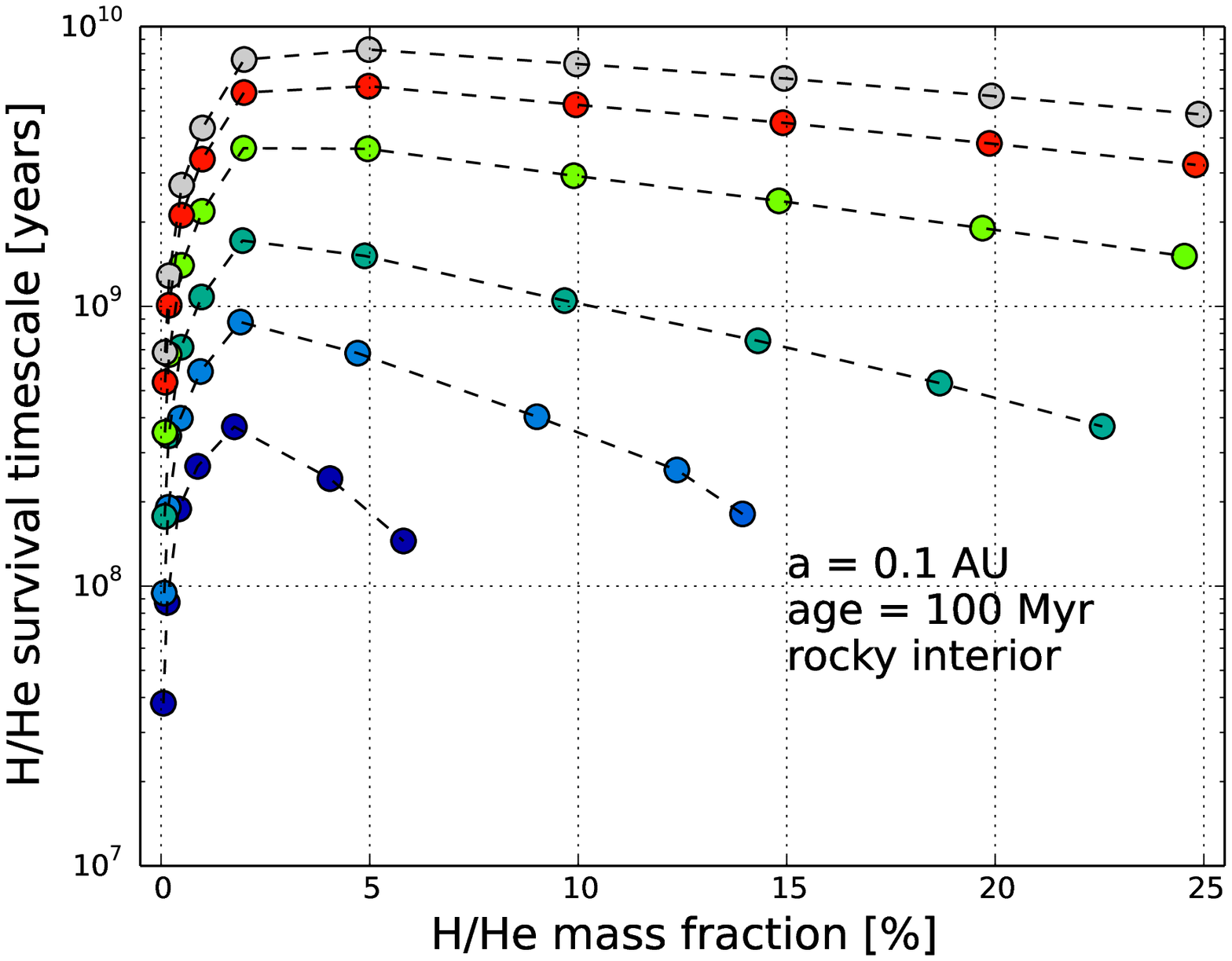}
\includegraphics[width=1.0\columnwidth]{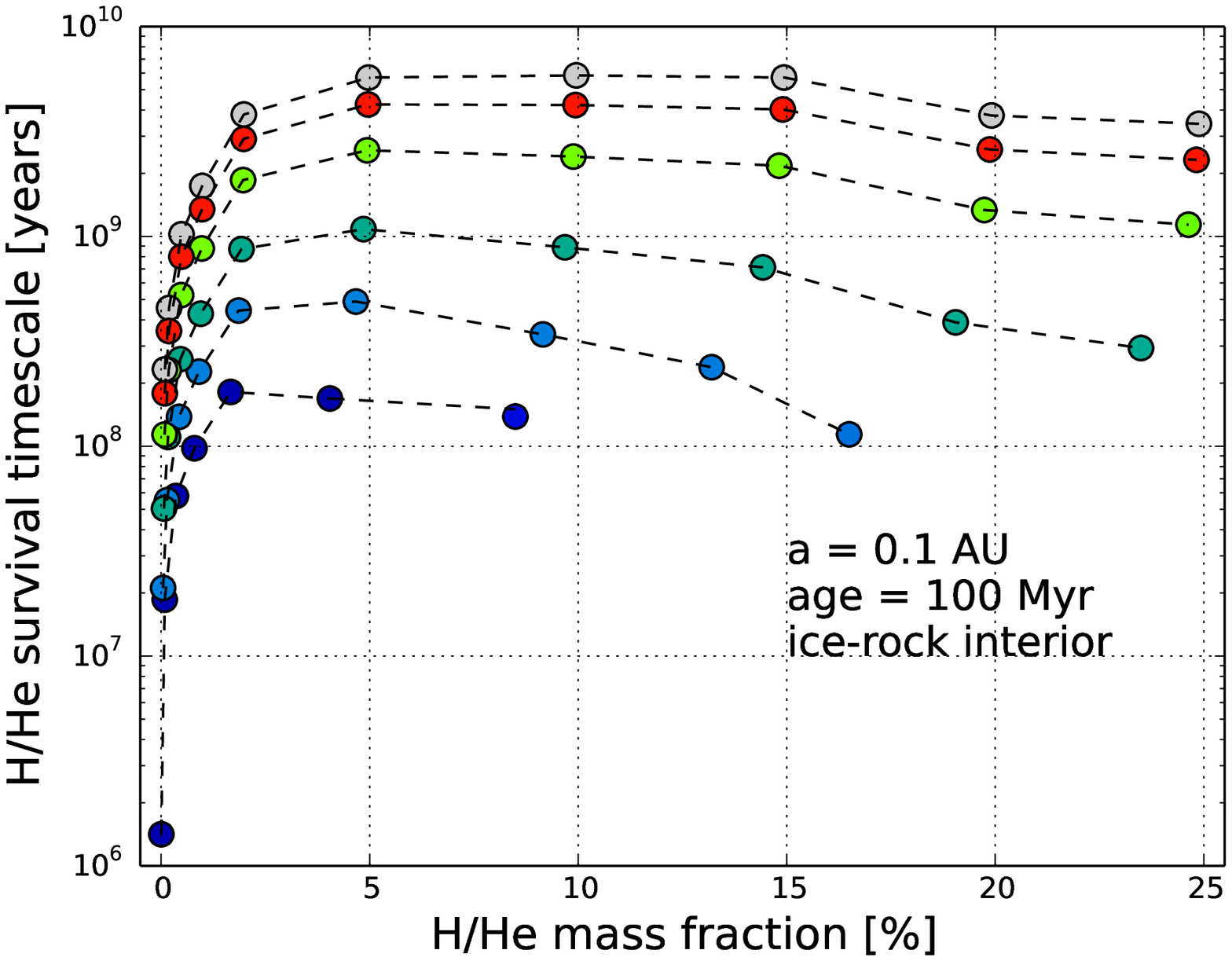}
\caption{\label{fig:tauvM}
Plot of instantaneous envelope mass-loss timescales, $\tau_{\rm env}$, as a function of H/He fraction and planet mass at 0.1 AU. These simulations have Earth-like rocky (top panel) or ice-rock (bottom panel) heavy element interiors, and solar metallicity envelopes. The initial planet masses for the scatter points from bottom to top are 4.0 (blue), 6.0 (light blue), 8.0 (teal), 13 (green), 17 (red), 20 (white) $M_{\oplus}$ respectively. The $\tau_{\rm env}$ values presented are instantaneous mass-loss rate at early times (100~Myr) when the stellar EUV is high. We see evidence of non-monotonic behavior of $\tau_{\rm env}$ with $f_{\rm env}$. At this instant, the most ``survivable" envelope fraction (at which $\tau_{\rm env}$ is maximized) is between 1 and 5\%.}
\end{center}  
\end{figure}

%Introducing the mass loss timescale
To more quantitatively examine the situation, it is illuminating to consider the planet envelope mass loss timescale, defined as $\tau_{\rm env} = M_{\rm env}/\dot M_{p}$ \citep[e.g.,][]{RogersEt2011ApJ, Batygin&Stevenson2013ApJ}. This quantity provides an instantaneous measure of how long a planet envelope could ``survive" at its current mass loss rate. For this experiment, we calculate $\tau_{\rm env}$ at early ages of 100 Myr. We present in Figure~\ref{fig:tauvM}, envelope mass loss timescales for planets at 0.1~AU.

%Quantitative results for envelope mass loss timescale

%result description
In our simulations of planets with rocky cores, the mass loss timescale (for specified planet mass) is maximized at an intermediate value of $f_{\rm env} = 1-2\%$.
At smaller envelope mass fractions ($f_{\rm env} = 0.05$ and 0.1\%), lifetimes are shorter because there is less envelope to lose. At higher envelope mass fractions above 5\%, planet radii increases the mass loss rate as the energy limited escape is sensitive to the cross-sectional radius (energy-limited $\propto R_{\rm p}^3$). % ; radiation-recombination $\propto R_{\rm p}^2$).
Furthermore, planets that loses a significant fraction of their total mass early on would likely completely evaporate before reaching ages of 1 Gyr \citep{LopezEt2012ApJ}. This may lead to convergent evolution behavior, where planets end up with similar final envelope masses fractions (near where the mass loss timescale is maximized) for a wide range of initial envelope mass fractions. This trend in our results also indicates that planets with $\sim 1\%$ H/He may ``linger" longer at the composition, compared to other values of $f_{\rm env}$. Once planets with high initial H/He fractions reaches this ${\sim}1\%$ value, they typically remain with similar values of envelope fraction for $\ga~5$ Gyr.

%Dependence on planet mass 
Upon closer examination of the top panel in Figure~\ref{fig:tauvM}, the planet envelope mass fraction at which the mass loss timescale is highest increases with planet mass. Below $12~M_{\oplus}$, the simulated planets with 1\% H/He mass fraction maximize $\tau_{\rm env}$. Above $12~M_{\oplus}$, the value of $f_{\rm env}$ that maximizes $\tau_{\rm env}$ switches to 5-6\%. 
This hints that ``fixed point" compositions to which planets converge may depend on the planet mass, potentially providing an observational diagnostic. %TODO: Think more about if we can make predictions based on these trends.

%Dependence on Orbital separation - may make sense to move near discussion of icy interiors
Recall, though, that these $\tau_{\rm env}$ values correspond to the stellar fluxes at 0.1 AU. At further out orbital distances ($\ga 0.1$ AU), the entire distribution shifts upward (toward higher $\tau_{\rm env}$). Conversely, at a closer-in distance, the envelope mass loss timescales decreases across all values of planet masses, largely preserving the general shape of the distribution.

%Comparing Core Comp.
For planets with lower density ice-rich cores (lower panel, Figure~\ref{fig:tauvM}), we see a distinct upward shift in the value of $f_{\rm env}$ that maximizes $\tau_{\rm env}$ compared to planets with higher-density rocky cores. 
The icy interior models do not generate the same 1\% signature but instead favor a higher set of ``most survivable" envelope mass fraction values (namely in the ${\gtrsim} 5\%$ range for planets $\gtrsim 5~M_{\oplus}$). This shift is a consequence of how a lower density heavy element interior distorts the planet $R_p-f_{\rm env}$ relation. 
At masses below ${\sim} 13~M_{\oplus}$, the ice-rock interior planets with $5-15\%$ by mass H/He envelopes all have similar trends and near maximal values of $\tau_{\rm env}$. At yet higher planet masses (above ${\sim} 13~M_{\oplus}$), the ${\sim} 15\%$ H/He composition takes over as the composition maximizing $\tau_{\rm env}$. Compared with the rocky-core models, this take-over point occurs at a much lower planet mass.

\begin{figure*}[t] %different options for where to place figure
\begin{center}
\includegraphics[width=2.0\columnwidth]{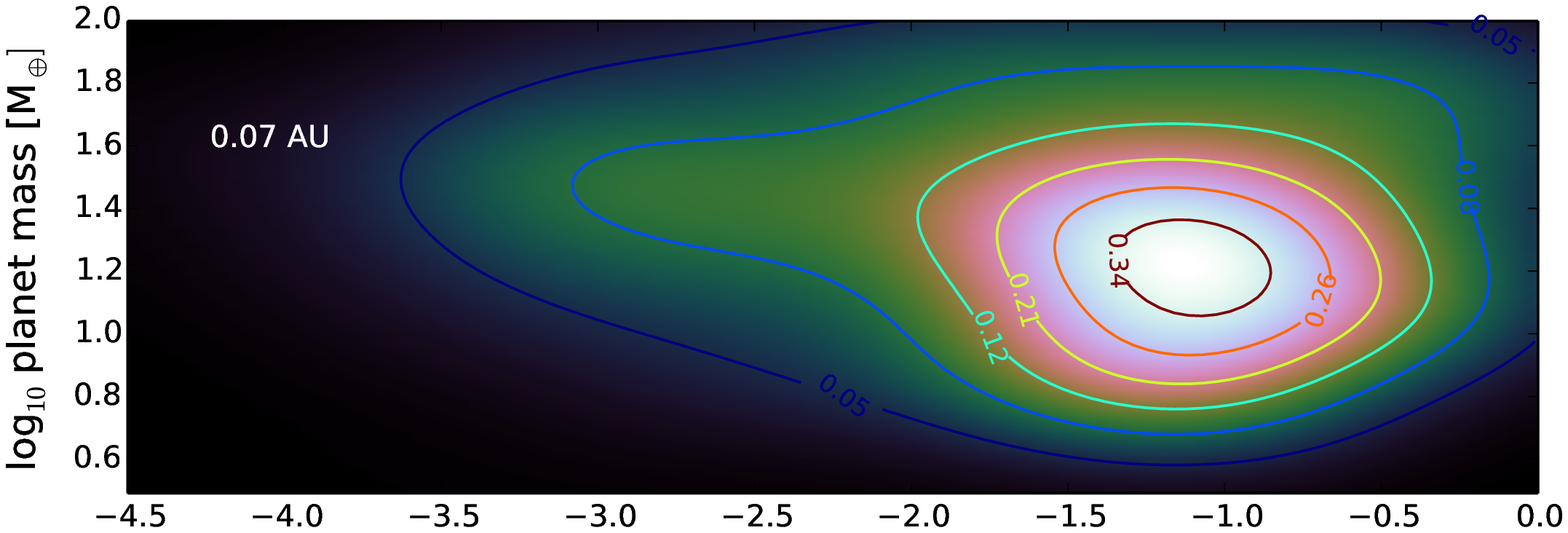}
\includegraphics[width=2.0\columnwidth]{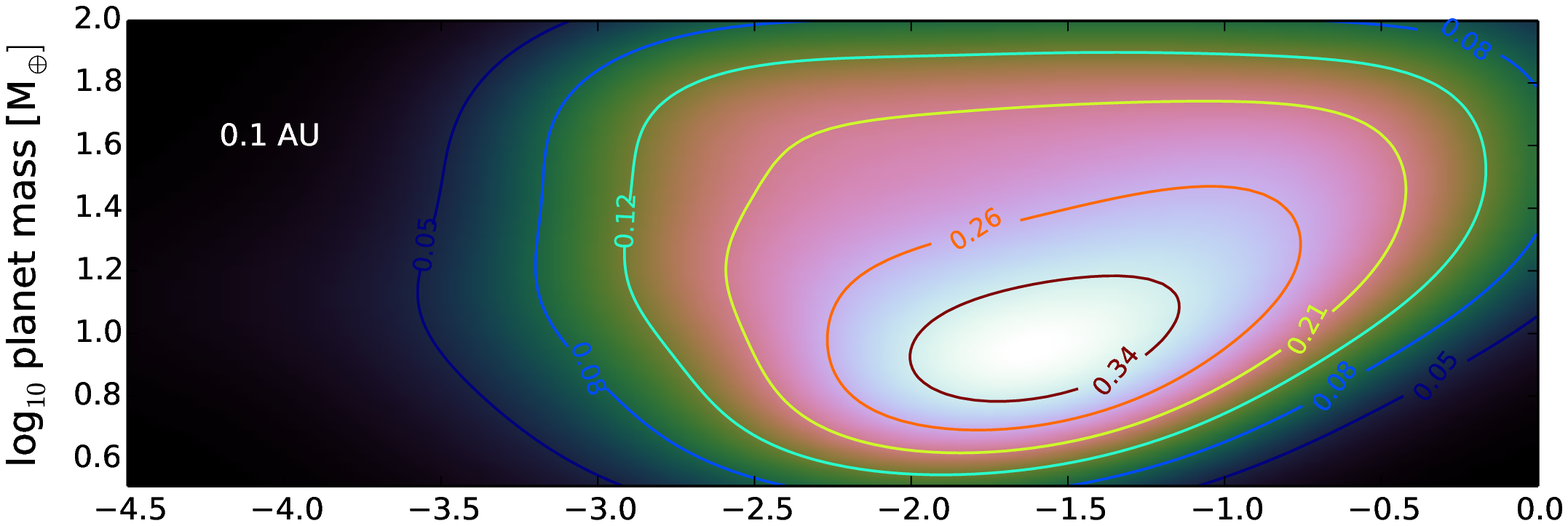}
\includegraphics[width=2.0\columnwidth]{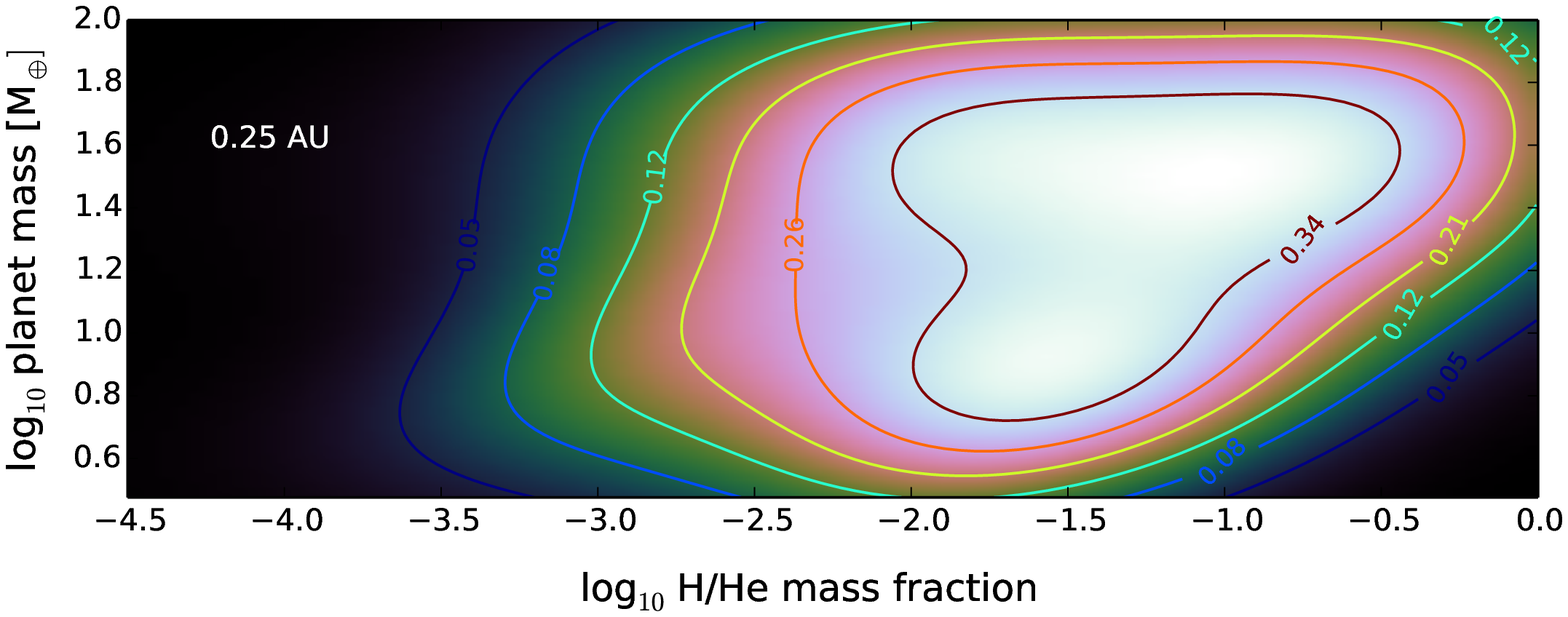}
\caption{\label{fig:2dcont}
Final mass-composition distribution of simulated mass-losing planets at 4.5 Gyr (top panel: 0.07 AU, middle panel: 0.1 AU, bottom panel 0.25 AU). The distribution of initial planet properties were chosen to be uniformly distributed in $\log M_p$-$\log f_{\rm env}$ space, with initial masses ranging from $1~M_{\oplus}$ to $100~M_{\oplus}$, and initial $f_{\rm env,0}$ from 0.001 to 0.95. We carried out calculations for 1200 simulated planets with randomly generated initial conditions for 4.5~Gyr. The final mass-composition distribution obtained from Gaussian kernel density estimation, is indicated by color-shading and contours, and represents a measure of the transfer function (at 4.5 Gyr), marginalized over $f_{\rm env,0}$.}
\end{center}
\end{figure*}

\begin{figure}[h] %different options for where to place figure
\begin{center}
\includegraphics[width=1.0\columnwidth]{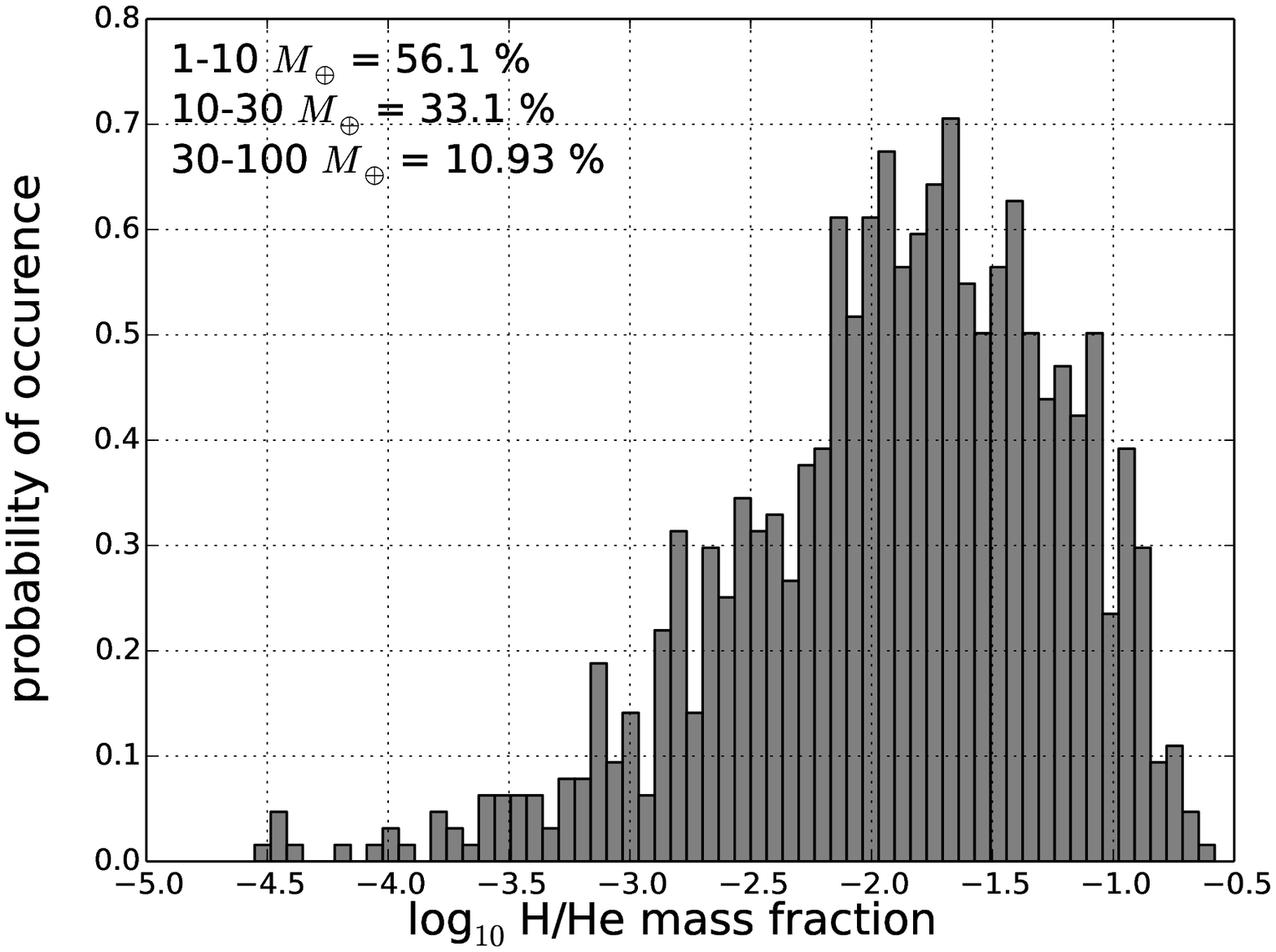}
\caption{\label{fig:histfit}
Synthetic planet probability-composition distribution at 4.5~Gyr and 0.1 AU. The initial $f_{\rm env,0}$ distribution assumed is flat in $\log f_{\rm env,0}$ (between $f_{\rm env,0}$ of 0.001 to 0.90). The planet mass distribution is sampled from the RV mass distribution of \citet{HowardEt2010Science}; the proportion of simulated planets within each mass bin is listed in the upper left corner of the plot. This 1D histogram is effectively Figure~\ref{fig:2dcont}, marginalized over planet mass, weighted by the California Planet Search radial velocity mass distribution.}
\end{center}
\end{figure}

A major caveat in this study of $\tau_{\rm env}$ is that even if the H/He envelopes mass fractions of 1-5\% surrounding rocky cores maximize the mass loss timescale at 100~Myr, we have not yet established whether this could have a substantial effect on the planet population, specifically whether this effect could plausibly reproduce the H/He mass fraction distribution noted by \citet{Wolfgang+Lopez2015ApJ}.
We calculated $\tau_{\rm env}$ by taking instantaneous mass-loss rates to predict planet survival lifetimes. Are the differences in $\tau_{\rm env}$ sufficiently large relative to the duration of active mass loss to leave an observable imprint on the planet population? To what extent would the planet composition distribution be altered over the course of Gyr timespans? In the next section, we turn to evaluating the evolution in the H/He inventory of a larger synthetic population of planets.

\subsection{Sampling a Simulated Planet Population}
\label{subsec:hist}

%Intro
In the previous section, we illustrated the effect of planet mass loss on the $f_{\rm env}$ distribution using instantaneous diagnostics at early times. We now synthesize all the implementations done so far to test the cumulative ramifications on planets in the present day. The planet population observed today (e.g., by {\it Kepler}) is a combination of both the initial outcomes from planet formation, and the subsequent effect of planet evolution. In this section, focus solely on the latter process, isolating how planet evolution would affect an arbitrary initial planet composition distribution. 

%Mathematical Formalism
We borrow the language of observational cosmology to express the effects of planet mass-loss evolution on the distribution, ${g}\left(M_{\rm core}, f_{\rm env}, a, t\right)$ of planet masses and compositions, at orbital separation, $a$, and time, $t$. The transfer function $T\left(M_{\rm core}, f_{\rm env}, f_{\rm env,0}, a, t\right)$ encapsulates how the initial planet mass-composition distribution output by the planet formation process, ${g}\left(M_{\rm core}, f_{\rm env}, a, t=0\right)$, will be modified over time

%\begin{equation}
\begin{multline*}
{g}\left(M_p, f_{\rm env}, a, t\right)= \\ 
\int T\left(M_{\rm core}, f_{\rm env}, f_{\rm env,0}, a, t\right){g}\left(M_{\rm core}, f_{\rm env, 0}, a,  t=0\right)df_{\rm env,0}.
\end{multline*}
\label{eq:TransFunc}
%\end{equation}

\noindent Planet mass loss causes planets to evolve towards lower envelope mass fractions (and lower total masses) over time.

%To illustrate the qualitative features of the planet evolution transfer function, we show planets from log 2 to log $50~M_{\oplus}$ and $f_{\rm env} =$ log 0.001 to log 0.90 at a fixed orbital distance of 0.1 AU. 
 
%Broad features of 2D histogram
We present in Figure~\ref{fig:2dcont} a snapshot in time (at 4.5~Gyr age) and orbital separations at 0.1, 0.07 and 0.25 AU of the planet evolution transfer function, marginalized over $f_{\rm env,0}$.  
The distribution of initial planet properties, ${g}\left(M_{\rm core}, f_{\rm env, 0}, a,  t=0\right)$, were chosen to be uniformly distributed in $\log M_p$-$\log f_{\rm env}$ space; any peaks and deserts in Figure~\ref{fig:2dcont} are due solely to the subsequent consequences of mass-loss evolution. We considered initial masses ranging from $1~M_{\oplus}$ to $100~M_{\oplus}$, and initial $f_{\rm env,0}$ from 0.001 to 0.90. For each orbital separation, we evolved 1200 planets with randomly generated conditions for 4.5~Gyr. Figure~\ref{fig:2dcont} presents final mass-composition distribution of the simulated planets, which effectively represent the distortion of the initial $\log M_p-\log f_{\rm env}$ distribution due to planet evolution.

%where are the peaks and deserts? 
We analyze some observations of the 0.1 AU diagram (large panel in Figure~\ref{fig:2dcont}). There is an over-density in the planet mass-composition distribution at envelope fractions of $\approx\ 10^{-2.0}$ to  $10^{-1.5}$ and planet masses $\la 20~M_{\oplus}$. Planets residing in this regime corresponds to those with compositions close to the most ``optimal" longest-lived envelope fractions (with highest $\tau_{\rm env}$).
This is the manifestation over the cumulative history of the planets of the non-monotonic behavior of $\tau_{\rm env}$ with $f_{\rm env}$ highlighted in Section~\ref{subsec:tau_env}). 

The 0.1 AU diagram in Figure~\ref{fig:2dcont} also displays a paucity of planets with masses $\la 17~M_{\oplus}$ surviving at 4.5~Gyr with $f_{\rm env} \la 10^{-3}\%$. This is due to the short envelope mass loss timescales in this regime (because there is little H/He to lose). 

For planets with masses above ${\sim} 16~M_{\oplus}$ ($10^{1.2}$), the ``distortion" of the original distribution ${g}\left(M_p, f_{\rm env, 0}, a, t=0\right)$ is less significant compared to the low mass regime and even less so for further orbital separations (compare the degree of distortion for the 0.07 AU and the 0.25 AU panels). Though there is an apparent dearth of planets with $f_{\rm env} \la 0.001$, this desert arises as a pure artifact due to the way in which we set the lower bound initial compositions ($f_{\rm env} \geq 0.001$). These results reflect the fact that planets in this regime are less likely to lose mass at this specific orbital distance, $F_{\rm p} \approx 100~F_{\oplus}$.

%Lower right-hand corner
Planet interior structure and evolution may also lead to low mass ($M_p \la 10~M_{\oplus}$) and high envelope mass fractions ($f_{\rm env} \ga 10\%$) being rare. This is visible in Figure~\ref{fig:2dcont} as an underdensity in the lower right-hand corner. This desert arises from the combined consequence of the massive mass-loss rates due to large planet cross-sectional area as well as the instability of MESA computations at even greater sizes (or $f_{\rm env}$). In the latter cases, planets in this regimes can be created but unable to be evolved past ages of ${\sim} 5-10$ Myr. Simulations with extremely low mean densities such as these cause the H/He envelopes to be ``unbounded" over a very short period of time (even without mass loss). The low-$M_p$-high-$f_{\rm env}$ boundary in the lower right corner of Figure~\ref{fig:2dcont} corresponds to the same boundary in the upper left corner of Figure~\ref{fig:rvm}.

Finally, we recover the convergent behavior of evolution tracks, in which points (if imagined as vector fields) move toward the lower left, along a constant core mass curve (not shown in figure). Particularly, a range of initial \% H/He values from 1 to 10\% ended up with a similar composition at 4.5~Gyr ($f_{\rm env}{\sim} 0.8\%$).

Ultimately, the transfer function (Figure~\ref{fig:2dcont}) is convolved with the initial distribution of planet properties (immediately after formation) to yield the current planet mass-composition distribution observed today. As a proof of concept, we sample from the mass distribution of radial velocity planets from the California Planet Search reported by \citet{HowardEt2010Science} to define a planet mass distribution (still assuming a flat distribution of initial $\log{f_{\rm env}}$). The resulting composition distribution of planets at 4.5~Gyr is shown in Figure~\ref{fig:histfit}.  
%In Figure~\ref{fig:histfit}, we show the composition distribution weighted from our simulations using the planet mass occurrence distribution reported by \citet{HowardEt2010Science} (instead of having the distribution being uniform across the entire mass range).
With this more realistic initial planet mass distribution, we can see a distinct peak of simulated planet occurrence at about 1\% H/He mass fraction. This result suggests that evolution via evaporation may partly explain the compositional distribution of sub-Neptune sized planets found by \citet{Wolfgang+Lopez2015ApJ}. 

In the past two sections, we have shown that photo-evaporation effects may, in certain regimes, lead to a favored envelope mass fraction where the envelope mass-loss timescale is optimized. However, we have focused only on evolutionary processes, encapsulated in the transfer function. The eventuality of planet evolution is also strongly shaped by the choices of initial masses and compositions. While the initial planet distribution we assumed here is an oversimplification, it nonetheless provides an encouraging demonstration that evolutionary processes play a large role in sculpting the planets observed today.

\section{Discussion}
\label{sec:discussion} 

\subsection{Insights into the {\it Kepler} Planet Population}

Planet interior structure and evolution calculations can provide insights into the observed distribution of {\it Kepler} planet properties. Several planet modelers have already noted features that photo-evaporation could produce in the radius-flux distribution of close-in planets, specifically a declining occurrence of sub-Neptune-size planets with increasing irradiation as vulnerable low-density planets lose their envelopes \citep[e.g.,][]{LeCavelierDesEtangs2007A&A, LopezEt2012ApJ}, and an ``occurrence valley" in the planet distribution between $\sim1.5~R_{\oplus}$ and $\sim2.5~R_{\oplus}$ between the populations of planets that have retained their volatile envelopes and the population of remnant evaporated cores \citep[e.g.,][]{Owen&Wu2013ApJ, Lopez&Fortney2013ApJ, MordasiniEt2012A&A, JinEt2014ApJ}. Our planets evolution simulations with MESA show good agreement (as expected) with these previously reported trends in the dividing line between complete and incomplete evaporation.

We have proposed yet another potential observable signature of evaporation in the close-in planet population, in Sections~\ref{subsec:tau_env} and \ref{subsec:hist}. Our simulations have hinted that the typical $\sim 1\%$ H/He mass fraction inferred from the {\it Kepler} radius distribution by \citet{Wolfgang+Lopez2015ApJ} could potentially be due to convergent evolution produced by evaporative planet mass loss. This feature appears among the population of planets that suffer significant but incomplete evaporation. With a toy model for the initial planet mass-composition distribution, we have provided an illustrative proof of concept that this effect can have an observable influence sculpting the eventual $f_{\rm env}$ distribution at Gyr ages. Further work is needed to more fully quantify the effect of convergent photo-evaporative evolution on the observed {\it Kepler} population, including  models incorporating a more sophisticated treatment of planet mass-loss, a fully de-biased joint radius-period distribution of Kepler planet candidates, and robust statistical methods to compare models and observations.
%Predictions would be great. Need to think more about this. 

\begin{figure*}[t]
\centering
  \begin{tabular}{@{}cccc@{}}
  \includegraphics[width=1.0\columnwidth]{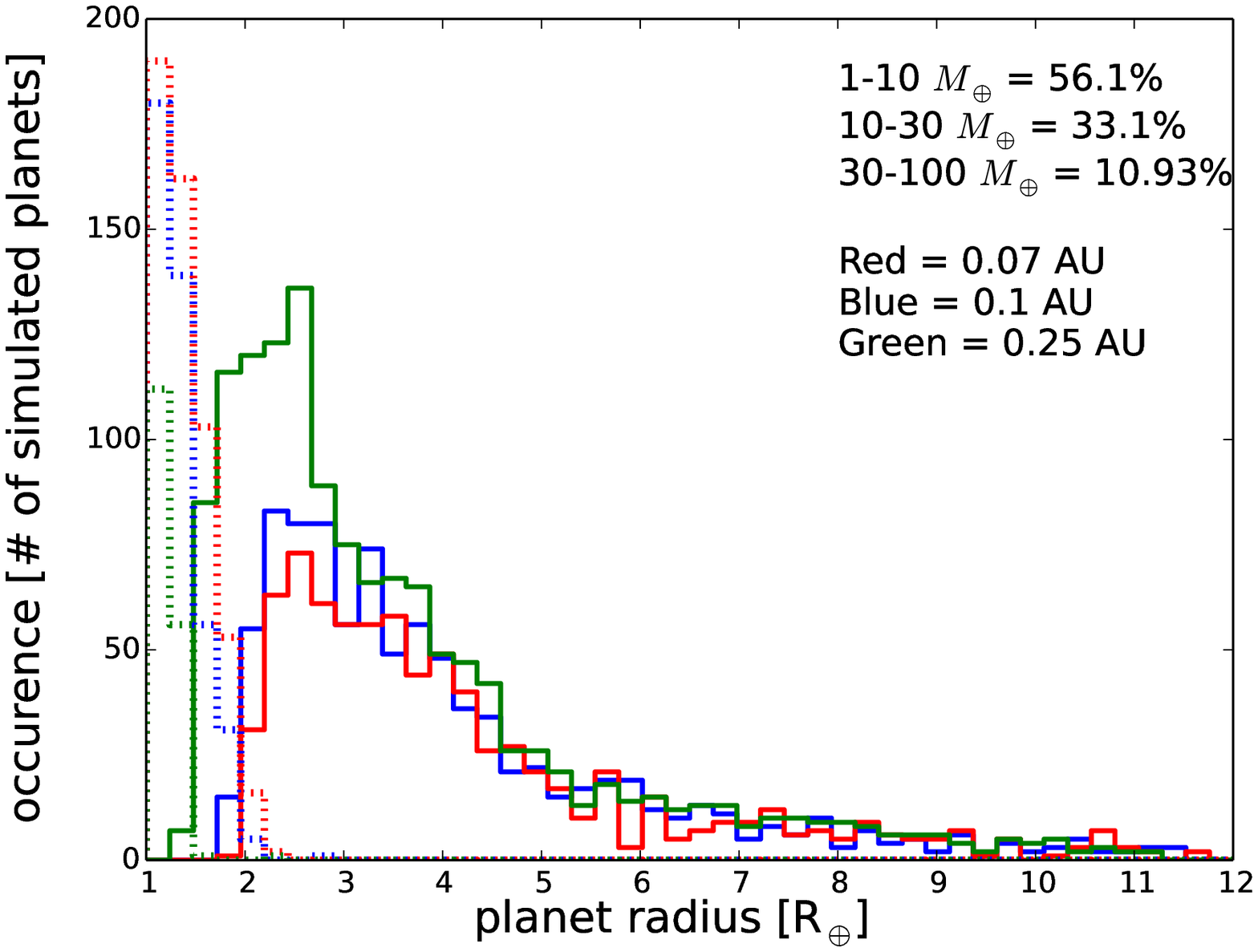}&
\includegraphics[width=1.0\columnwidth]{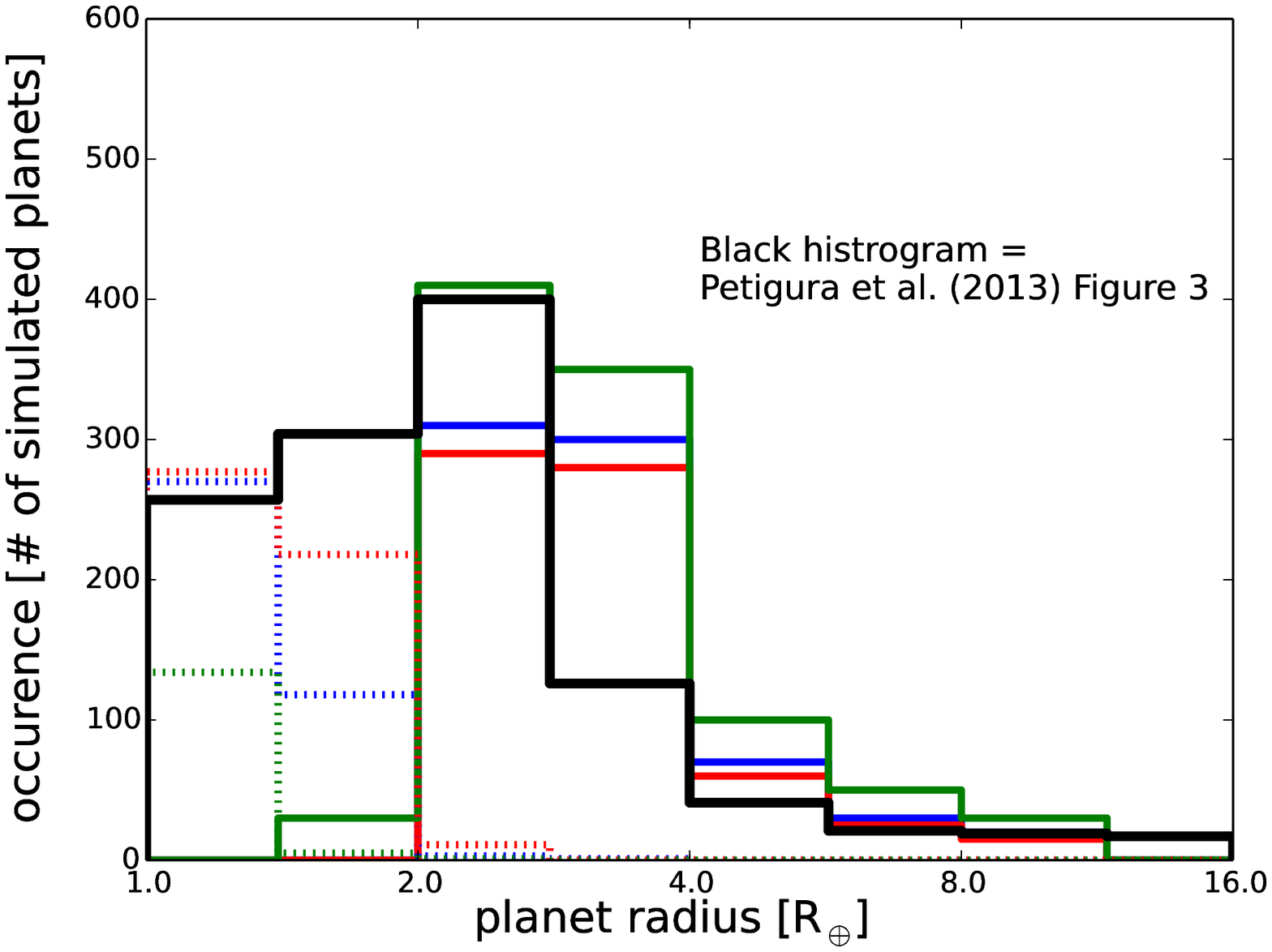}\\
  \end{tabular}
  \caption{\label{fig:discussion}
Synthetic planet radius distributions at 4.5~Gyr on a linear radius scale (left) and logarithmic radius scale (right). The histogram colors indicate different orbital separations: 0.07~AU (red), 0.10~AU (blue) and 0.25~AU (green). A total of 1200 planets were simulated at each orbital separation. The solid lines represent planet models that have retained some H/He in their envelopes, while dotted lines correspond to evaporated planet cores. The blue histogram at 0.10~AU corresponds to the same suite of simulations presented in Figure~\ref{fig:histfit}. The bottom panel is plotted to ease comparison with the {\it Kepler} radius occurrence rates from \citet{PetiguraEt2013PNAS} (black step-curve).}
\end{figure*}

Ultimately, {\it Kepler} measured planet radii (transit depths) and it is to the distribution of planet radii that any model should be compared. %Indeed, the typical $\sim 1\%$ H/He mass fraction inferred by \citet{Wolfgang+Lopez2015ApJ} stems from the {\it Kepler} radius distribution. 
In Figure~\ref{fig:discussion} we present the distribution of planet radii associated with the synthetic planet population at 0.1~AU in Figure~\ref{fig:histfit} along with synthetic planet radius distributions at 0.07~AU and 0.25~AU, generated following an identical approach. Our synthetic population shows the occurrence valley between evaporated cores and volatile rich planets (between 1 and $2~R_{\oplus}$) that was found in previous theoretical works. The planet radius corresponding to the valley decreases with increasing orbital separation. Like in the observed {\it Kepler} radius distribution from \citet{PetiguraEt2013PNAS}, our synthetic radius distributions display a occurrence peak in the 2 to $2.83~R_{\oplus}$ bin (when logarithmic radius bins are employed). On the other hand, \citet{PetiguraEt2013PNAS} see a much steeper decline in planet occurrence between the 2-$2.8~R_{\oplus}$ and 2.8-$4~R_{\oplus}$ bins than is present in our synthetic population. We emphasize however, that we assumed one of the simplest and broadest initial $M_p-f_{\rm env}$ distributions possible (flat in $\log f_{\rm env}$, and no correlation between initial $M_p$ and $f_{\rm env}$); we made no attempt to tune the initial mass-composition distribution to fit the observed {\it Kepler} radius distribution. The steeper observed decline around 3~$R_{\oplus}$ could be a consequence of a narrower distribution of initial compositions output from formation, evidence for a sub-population of water-rich planets, and/or correlations between planet initial planet mass and $f_{\rm env}$. Indeed, planet formation models predict that correlations should exist between envelope mass and core mass \citep{Bodenheimer&Lissauer2014ApJ, MordasiniEt2014A&A, Lee&Chiang2015ApJ}. Further statistical studies linking {\it Kepler} data to models are warranted to better disentangle the effects of evolution and the outcomes of formation.

\subsection{Model Extension Opportunities}
\label{sec:Extensions}

We extended the MESA stellar evolution code to simulate H/He envelopes surrounding low-mass planets in 1D. The physics that we incorporated into MESA is not necessarily new; rather, we followed common assumptions and approaches employed in closed-source (proprietary) 1D planet evolution codes that are in use in the field (e.g. \citealt{ValenciaEt2010A&A}, \citealt{Lopez+Fortney2014ApJ}, \citealt{Kurokawa&Nakamoto2014ApJ}, and \citealt{Howe&Burrows2015}). With our adaptations to MESA, there now exists a publicly available open-source code to simulate H/He planets down to a few Earth masses. 

True planets are surely more complicated than the vanilla spherically symmetric, homogeneously layered scenarios simulated in this paper \citep[and in prior works][]{ValenciaEt2010A&A,Lopez+Fortney2014ApJ,Kurokawa&Nakamoto2014ApJ,Howe&Burrows2015}. These 1D simulations are, nonetheless, the current workhorses of exoplanet evolution studies, and help to provide context to more complex and computationally intensive models.

There are a number of physical processes that could be added to MESA in future work, to improve upon its capabilities to model planets. 

These include, a treatment of composition gradients within planet interiors (e.g. double diffusive convection \citealt{NettelmannEt2015}), using self-consistent model planet atmosphere grids to set the outer boundary conditions and planet cooling rates, \citep[as in, e.g.,][]{FortneyEt2007ApJ}, a more sophisticated treatment of atmospheric escape (including energy-limited scaling laws \citep{SalzEt2015arXiv} and MHD effects), and finally the addition of a water EOS to facilitate the simulation of high mean molecular weight planetary envelopes.

It is our hope that the adaptations to MESA presented in this paper will help to provide the baseline groundwork for future applications of MESA to low-mass planets.

\section{Summary \& Conclusions} 
\label{sec:sum}

%Summary
We summarize briefly below, the main outcomes and conclusions of this work.
\begin{itemize}
\item We implemented extensions to the MESA stellar evolution code that now permit MESA to simulate H/He envelopes surrounding planets down to a few Earth masses. These extensions include a thermo-physical model for planet heavy-element interiors, energy-limited and radiation-recombination limited mass loss, and an improved atmospheric boundary condition.
\item Coupled thermal and mass-loss evolution confirm that ultra low-density planets (with radii above $10~R_{\oplus}$ and masses below $30~M_{\oplus}$) can plausibly survive for multiple Gyr timescales, though over a narrower range of parameter space that predicted by instantaneous mass loss timescale criteria \citep{Batygin&Stevenson2013ApJ}.
\item For mass-losing planets even at close orbital distances $\la 0.1$ AU, radius is a proxy for the planet's \emph{current} composition.
\item Planet radii typically show very little hysteresis. The systematic error introduced by applying planet isochrons calculated neglecting mass loss to define a mapping from planet mass, composition, and age to radius for evaporating planets is typically small ($\lesssim 1\%$ in planet radius), with the exception of models with very high envelope fractions.
\item Planet envelope mass loss timescales, $\tau_{\rm env}$ vary non-monotonically with $f_{\rm env}$ (at fixed planet mass). In our simulations of young (100~Myr) low-mass ($M_p\lesssim10~M_\oplus$) planets with rocky cores, $\tau_{\rm env}$ is maximized at $f_{\rm env}=1\%$ to 3\%. The resulting convergent evolution could potentially imprint itself on the close-in planet population as a preferred H/He mass fraction of $\sim1\%$ \citep[as inferred from the {\it Kepler} radius distribution by][]{Wolfgang+Lopez2015ApJ}
\end{itemize}

%Conclusion
With a succession of space-based exoplanet transit surveys on the horizon ({\it K2, TESS, CHEOPS} and {\it PLATO}) combined with improving resolution and stability of ground-based spectrographs (e.g., SPIRou, Keck SHREK, EXPRES, Carmenes, HPF, ESPRESSO, G-CLEF, HiJaK), our purview of exoplanetary systems is bound to expand vastly in the years to come. For this reason, there is a need for a fast yet robust series of modeling and computational schemes to complement observational measurements. Recent years has seen a near simultaneous rise in the use of the stellar evolution code MESA and planetary mass loss evolution studies. In this article, we provide a suite of basic planet models from which more complicated studies can be built; this is again eased by the open-source nature of MESA. Looking ahead, we see this wonderful evolution code acting as a complement to more sophisticated 3-D models to interpret the measurements of future space missions and exoplanetary surveys.

\acknowledgements
We thank Dr. James Owen for assisting us with the atmospheric boundary conditions and Professor Phil Arras for providing us with helpful templates that enhanced the efficiency of simulating MESA planets. We also thank Dr. Eric Lopez for great suggestions to our manuscript. H.C. acknowledges the Undergraduate Research Opportunities Program (UROP) at Boston University for funding this research while in residence at Caltech during the summer of 2014 and throughout the academic year. The authors thank Professors Philip Muirhead and Heather Knutson for facilitating the summer research. L.A.R. gratefully acknowledges support provided by NASA through Hubble Fellowship grant \#HF-51313 awarded by the Space Telescope Science Institute, which is operated by the Association of Universities for Research in Astronomy, Inc., for NASA, under contract NAS 5-26555. This work was performed in part under contract with the Jet Propulsion Laboratory (JPL) funded by NASA through the Sagan Fellowship Program executed by the NASA Exoplanet Science Institute. Most of the calculations have made use of H.C.'s MSI GE70 APACHE laptop, which was supported by his parents to pursue his interest in computational astrophysics and planetary science. Lastly, we express gratitude toward the MESA code creators$-$Dr. Bill Paxton, and Professors Lars Bildsten and Frank Timmes$-$without whom this project would not have been a possibility.

\bibliography{exoplanets}

\newpage
\appendix

%\section{appendix section}

\begin{deluxetable}{cccccccccccccccc}
%% Keep a portrait orientation

%% Over-ride the default font size
%% Use Default (12pt)

%% Use \tablewidth{?pt} to over-ride the default table width.
%% If you are unhappy with the default look at the end of the
%% *.log file to see what the default was set at before adjusting
%% this value.

%% This is the title of the table.
\tablecaption{Planetary Radii Table of Jovian and Sub-Jovian Gas Giants with 50\% Ice - 50\%Rock Cores}

%\tablenum{1}

\tablehead{\colhead{Age} & \colhead{Separation} & \colhead{Core Mass} & \colhead{0.0535} & \colhead{0.0881} & \colhead{0.115} & \colhead{0.242} & \colhead{0.406} & \colhead{0.676} & \colhead{1.0} & \colhead{1.46} & \colhead{2.44} & \colhead{4.07} & \colhead{6.78} & \colhead{11.31} & \colhead{} \\ 
\colhead{(Gyr)} & \colhead{(AU)} & \colhead{($M_{\oplus}$)} & \colhead{} & \colhead{} & \colhead{} & \colhead{} & \colhead{} & \colhead{} & \colhead{} & \colhead{} & \colhead{} & \colhead{} & \colhead{} & \colhead{} & \colhead{} } 

%% All data must appear between the \startdata and \enddata commands
\startdata
0.3 & 0.02 & 0.0 & nan & nan & nan & nan & 15.32 & 14.47 & 14.09 & 14.0 & 14.02 & 14.0 & 13.87 & 13.55 &  \\
0.3 & 0.02 & 10.0 & 13.17 & 15.37 & 15.26 & 14.57 & 14.22 & 13.87 & 13.79 & 13.83 & 13.95 & 14.04 & 13.98 & 13.69 &  \\
0.3 & 0.02 & 25.0 & nan & 5.12 & 8.28 & 11.95 & 12.79 & 13.15 & 13.33 & 13.54 & 13.8 & 13.98 & 13.98 & 13.74 &  \\
0.3 & 0.02 & 50.0 & nan & nan & nan & 8.18 & 10.7 & 11.98 & 12.6 & 13.04 & 13.51 & 13.82 & 13.92 & 13.75 &  \\
0.3 & 0.02 & 100.0 & nan & nan & nan & nan & 6.73 & 9.8 & 11.14 & 12.05 & 12.91 & 13.47 & 13.73 & 13.68 &  \\
0.3 & 0.045 & 0.0 & nan & nan & nan & nan & 13.56 & 13.28 & 13.14 & 13.14 & 13.19 & 13.16 & 12.94 & 12.53 &  \\
0.3 & 0.045 & 10.0 & 9.23 & 11.34 & 11.8 & 12.45 & 12.71 & 12.82 & 12.87 & 12.98 & 13.14 & 13.2 & 13.05 & 12.69 &  \\
0.3 & 0.045 & 25.0 & nan & 4.62 & 7.17 & 10.51 & 11.57 & 12.16 & 12.47 & 12.73 & 13.01 & 13.15 & 13.07 & 12.76 &  \\
0.3 & 0.045 & 50.0 & nan & nan & nan & 7.5 & 9.84 & 11.14 & 11.81 & 12.28 & 12.76 & 13.02 & 13.03 & 12.79 &  \\
0.3 & 0.045 & 100.0 & nan & nan & nan & nan & 6.4 & 9.22 & 10.51 & 11.39 & 12.22 & 12.71 & 12.88 & 12.75 &  \\
0.3 & 0.1 & 0.0 & nan & nan & nan & nan & 13.24 & 13.08 & 13.0 & 13.02 & 13.1 & 13.08 & 12.86 & 12.43 &  \\
0.3 & 0.1 & 10.0 & 8.25 & 10.41 & 11.02 & 12.02 & 12.42 & 12.63 & 12.73 & 12.87 & 13.05 & 13.11 & 12.96 & 12.58 &  \\
0.3 & 0.1 & 25.0 & nan & 4.47 & 6.88 & 10.2 & 11.33 & 12.0 & 12.34 & 12.62 & 12.92 & 13.07 & 12.98 & 12.66 &  \\
0.3 & 0.1 & 50.0 & nan & nan & nan & 7.34 & 9.67 & 11.0 & 11.68 & 12.18 & 12.67 & 12.94 & 12.95 & 12.69 &  \\
0.3 & 0.1 & 100.0 & nan & nan & nan & nan & 6.33 & 9.12 & 10.41 & 11.3 & 12.14 & 12.63 & 12.8 & 12.66 &  \\
0.3 & 1.0 & 0.0 & nan & nan & nan & nan & 12.44 & 12.53 & 12.58 & 12.68 & 12.86 & 12.94 & 12.82 & 12.41 &  \\
0.3 & 1.0 & 10.0 & 6.98 & 8.99 & 9.69 & 11.05 & 11.7 & 12.1 & 12.31 & 12.53 & 12.81 & 12.97 & 12.91 & 12.56 &  \\
0.3 & 1.0 & 25.0 & nan & 4.23 & 6.35 & 9.51 & 10.73 & 11.52 & 11.94 & 12.29 & 12.69 & 12.93 & 12.94 & 12.64 &  \\
0.3 & 1.0 & 50.0 & nan & nan & nan & 6.99 & 9.23 & 10.6 & 11.33 & 11.88 & 12.45 & 12.81 & 12.9 & 12.68 &  \\
0.3 & 1.0 & 100.0 & nan & nan & nan & nan & 6.15 & 8.84 & 10.13 & 11.05 & 11.94 & 12.51 & 12.76 & 12.64 &  \\
1.0 & 0.02 & 0.0 & nan & nan & nan & nan & 14.1 & 13.61 & 13.37 & 13.33 & 13.36 & 13.34 & 13.14 & 12.73 &  \\
1.0 & 0.02 & 10.0 & 11.24 & 13.1 & 13.22 & 13.19 & 13.18 & 13.12 & 13.11 & 13.18 & 13.31 & 13.36 & 13.22 & 12.86 &  \\
1.0 & 0.02 & 25.0 & nan & 4.83 & 7.59 & 11.0 & 11.95 & 12.44 & 12.69 & 12.91 & 13.18 & 13.31 & 13.24 & 12.93 &  \\
1.0 & 0.02 & 50.0 & nan & nan & nan & 7.72 & 10.1 & 11.37 & 12.0 & 12.45 & 12.91 & 13.18 & 13.2 & 12.96 &  \\
1.0 & 0.02 & 100.0 & nan & nan & nan & nan & 6.48 & 9.37 & 10.66 & 11.54 & 12.36 & 12.86 & 13.04 & 12.92 &  \\
1.0 & 0.045 & 0.0 & nan & nan & nan & nan & 12.76 & 12.68 & 12.62 & 12.65 & 12.72 & 12.68 & 12.44 & 11.99 &  \\
1.0 & 0.045 & 10.0 & 8.2 & 10.14 & 10.67 & 11.58 & 12.01 & 12.26 & 12.38 & 12.52 & 12.68 & 12.72 & 12.54 & 12.14 &  \\
1.0 & 0.045 & 25.0 & nan & 4.42 & 6.7 & 9.87 & 10.99 & 11.66 & 12.0 & 12.28 & 12.57 & 12.69 & 12.58 & 12.23 &  \\
1.0 & 0.045 & 50.0 & nan & nan & nan & 7.16 & 9.41 & 10.71 & 11.38 & 11.86 & 12.33 & 12.57 & 12.55 & 12.27 &  \\
1.0 & 0.045 & 100.0 & nan & nan & nan & nan & 6.21 & 8.91 & 10.16 & 11.02 & 11.82 & 12.28 & 12.42 & 12.25 &  \\
1.0 & 0.1 & 0.0 & nan & nan & nan & nan & 12.45 & 12.49 & 12.48 & 12.54 & 12.63 & 12.6 & 12.37 & 11.91 &  \\
1.0 & 0.1 & 10.0 & 7.39 & 9.37 & 10.0 & 11.18 & 11.74 & 12.08 & 12.24 & 12.41 & 12.59 & 12.64 & 12.47 & 12.06 &  \\
1.0 & 0.1 & 25.0 & nan & 4.27 & 6.44 & 9.59 & 10.76 & 11.49 & 11.87 & 12.17 & 12.48 & 12.61 & 12.5 & 12.15 &  \\
1.0 & 0.1 & 50.0 & nan & nan & nan & 7.01 & 9.24 & 10.57 & 11.26 & 11.76 & 12.25 & 12.5 & 12.48 & 12.2 &  \\
1.0 & 0.1 & 100.0 & nan & nan & nan & nan & 6.14 & 8.81 & 10.07 & 10.93 & 11.74 & 12.21 & 12.35 & 12.18 &  \\
1.0 & 1.0 & 0.0 & nan & nan & nan & nan & 11.76 & 11.98 & 12.08 & 12.19 & 12.34 & 12.38 & 12.24 & 11.87 &  \\
1.0 & 1.0 & 10.0 & 6.41 & 8.27 & 8.97 & 10.39 & 11.11 & 11.6 & 11.85 & 12.06 & 12.3 & 12.41 & 12.33 & 12.02 &  \\
1.0 & 1.0 & 25.0 & nan & 4.09 & 6.03 & 9.01 & 10.23 & 11.06 & 11.5 & 11.84 & 12.19 & 12.38 & 12.37 & 12.11 &  \\
1.0 & 1.0 & 50.0 & nan & nan & nan & 6.72 & 8.86 & 10.2 & 10.93 & 11.45 & 11.97 & 12.28 & 12.35 & 12.15 &  \\
1.0 & 1.0 & 100.0 & nan & nan & nan & nan & 5.99 & 8.56 & 9.79 & 10.67 & 11.5 & 12.01 & 12.22 & 12.13 &  \\
4.5 & 0.02 & 0.0 & nan & nan & nan & nan & 12.86 & 12.72 & 12.64 & 12.65 & 12.71 & 12.68 & 12.44 & 11.99 &  \\
4.5 & 0.02 & 10.0 & 9.43 & 11.03 & 11.32 & 11.79 & 12.1 & 12.3 & 12.4 & 12.52 & 12.68 & 12.71 & 12.54 & 12.13 &  \\
4.5 & 0.02 & 25.0 & nan & 4.6 & 6.91 & 10.0 & 11.05 & 11.69 & 12.02 & 12.28 & 12.56 & 12.68 & 12.57 & 12.22 &  \\
4.5 & 0.02 & 50.0 & nan & nan & nan & 7.2 & 9.44 & 10.73 & 11.39 & 11.86 & 12.33 & 12.57 & 12.55 & 12.27 &  \\
4.5 & 0.02 & 100.0 & nan & nan & nan & nan & 6.2 & 8.9 & 10.16 & 11.02 & 11.82 & 12.28 & 12.41 & 12.24 &  \\
4.5 & 0.045 & 0.0 & nan & nan & nan & nan & 11.83 & 11.98 & 12.04 & 12.12 & 12.22 & 12.19 & 11.94 & 11.47 &  \\
4.5 & 0.045 & 10.0 & 7.24 & 8.98 & 9.54 & 10.62 & 11.2 & 11.61 & 11.83 & 12.01 & 12.2 & 12.23 & 12.05 & 11.64 &  \\
4.5 & 0.045 & 25.0 & nan & 4.23 & 6.24 & 9.16 & 10.3 & 11.07 & 11.48 & 11.79 & 12.09 & 12.21 & 12.1 & 11.74 &  \\
4.5 & 0.045 & 50.0 & nan & nan & nan & 6.77 & 8.89 & 10.2 & 10.9 & 11.4 & 11.88 & 12.12 & 12.09 & 11.8 &  \\
4.5 & 0.045 & 100.0 & nan & nan & nan & nan & 5.98 & 8.54 & 9.76 & 10.61 & 11.4 & 11.85 & 11.97 & 11.79 &  \\
4.5 & 0.1 & 0.0 & nan & nan & nan & nan & 11.57 & 11.81 & 11.91 & 12.02 & 12.14 & 12.11 & 11.88 & 11.41 &  \\
4.5 & 0.1 & 10.0 & 6.56 & 8.34 & 8.98 & 10.29 & 10.96 & 11.45 & 11.7 & 11.91 & 12.11 & 12.16 & 11.98 & 11.57 &  \\
4.5 & 0.1 & 25.0 & nan & 4.09 & 6.0 & 8.91 & 10.11 & 10.92 & 11.36 & 11.69 & 12.01 & 12.14 & 12.03 & 11.68 &  \\
4.5 & 0.1 & 50.0 & nan & nan & nan & 6.64 & 8.75 & 10.08 & 10.79 & 11.31 & 11.8 & 12.05 & 12.02 & 11.74 &  \\
4.5 & 0.1 & 100.0 & nan & nan & nan & nan & 5.92 & 8.45 & 9.68 & 10.53 & 11.33 & 11.78 & 11.91 & 11.73 &  \\
4.5 & 1.0 & 0.0 & nan & nan & nan & nan & 11.1 & 11.44 & 11.6 & 11.73 & 11.87 & 11.87 & 11.67 & 11.28 &  \\
4.5 & 1.0 & 10.0 & 5.86 & 7.6 & 8.29 & 9.77 & 10.53 & 11.1 & 11.4 & 11.62 & 11.84 & 11.91 & 11.77 & 11.44 &  \\
4.5 & 1.0 & 25.0 & nan & 3.93 & 5.72 & 8.53 & 9.75 & 10.61 & 11.08 & 11.42 & 11.75 & 11.9 & 11.82 & 11.54 &  \\
4.5 & 1.0 & 50.0 & nan & nan & nan & 6.45 & 8.49 & 9.82 & 10.53 & 11.06 & 11.55 & 11.81 & 11.82 & 11.6 &  \\
4.5 & 1.0 & 100.0 & nan & nan & nan & nan & 5.83 & 8.28 & 9.47 & 10.31 & 11.1 & 11.56 & 11.71 & 11.6 &  \\
\enddata

%% Include any \tablenotetext{key}{text}, \tablerefs{ref list},
%% or \tablecomments{text} between the \enddata and 
%% \end{deluxetable} commands

%% General table comment marker
\tablecomments{Top row are total mass values in $M_{\rm Jup}$. The planet radii values in $R_{\oplus}$. The specific mass and flux values were chosen to ease comparison with Fortney et al. (2007).}

%% No \tablerefs indicated
\label{tab:Fort07} 
\end{deluxetable}

%% there are and how to align them.
\begin{deluxetable}{cccccccccccccc}

%% Keep a portrait orientation

%% Over-ride the default font size
%% Use Default (12pt)

%% Use \tablewidth{?pt} to over-ride the default table width.
%% If you are unhappy with the default look at the end of the
%% *.log file to see what the default was set at before adjusting
%% this value.

%% This is the title of the table.
\tablecaption{Planetary Radii Table of Sub-Neptune Mass Planets with Earth-like Rocky Interiors at 10 Gyr}

%% This command over-rides LaTeX's natural table count
%% and replaces it with this number.  LaTeX will increment 
%% all other tables after this table based on this number
\tablenum{2}

%% The \tablehead gives provides the column headers.  It
%% is currently set up so that the column labels are on the
%% top line and the units surrounded by ()s are in the 
%% bottom line.  You may add more header information by writing
%% another line between these lines. For each column that requries
%% extra information be sure to include a \colhead{text} command
%% and remember to end any extra lines with \\ and include the 
%% correct number of &s.
\tablehead{\colhead{Stellar Flux} & \colhead{Planet Mass} & \colhead{0.01\%} & \colhead{0.02\%} & \colhead{0.05\%} & \colhead{0.1\%} & \colhead{0.2\%} & \colhead{0.5\%} & \colhead{1.0\%} & \colhead{2.0\%} & \colhead{5.0\%} & \colhead{10\%} & \colhead{20\%} & \colhead{} \\ 
\colhead{($F_\oplus$)} & \colhead{($M_\oplus$)} & \colhead{} & \colhead{} & \colhead{} & \colhead{} & \colhead{} & \colhead{} & \colhead{} & \colhead{} & \colhead{} & \colhead{} & \colhead{} & \colhead{} } 

%% All data must appear between the \startdata and \enddata commands
\startdata
0.1 & 1.0 & 1.03 & 1.05 & 1.08 & 1.11 & 1.15 & 1.23 & 1.34 & 1.5 & 1.87 & 2.38 & 3.27 &  \\
0.1 & 1.5 & 1.15 & 1.17 & 1.19 & 1.22 & 1.26 & 1.34 & 1.44 & 1.6 & 1.96 & 2.43 & 3.25 &  \\
0.1 & 2.0 & 1.24 & 1.26 & 1.28 & 1.31 & 1.35 & 1.43 & 1.53 & 1.68 & 2.03 & 2.5 & 3.29 &  \\
0.1 & 2.4 & 1.3 & 1.32 & 1.34 & 1.37 & 1.41 & 1.49 & 1.59 & 1.74 & 2.09 & 2.55 & 3.33 &  \\
0.1 & 3.6 & 1.45 & 1.46 & 1.49 & 1.51 & 1.55 & 1.63 & 1.73 & 1.88 & 2.23 & 2.69 & 3.45 &  \\
0.1 & 5.5 & 1.62 & 1.63 & 1.65 & 1.68 & 1.72 & 1.8 & 1.9 & 2.05 & 2.32 & 2.87 & 3.62 &  \\
0.1 & 8.5 & 1.8 & 1.81 & 1.84 & 1.86 & 1.9 & 1.98 & 2.08 & 2.2 & 2.59 & 3.08 & 3.83 &  \\
0.1 & 13.0 & 2.0 & 2.01 & 2.03 & 2.06 & 2.09 & 2.18 & 2.24 & 2.51 & 2.83 & 3.31 & 4.07 &  \\
0.1 & 20.0 & 2.2 & 2.21 & 2.24 & 2.26 & 2.3 & 2.37 & 2.43 & nan & nan & nan & nan &  \\
10.0 & 1.0 & 1.17 & 1.19 & 1.24 & 1.29 & 1.35 & 1.47 & 1.62 & 1.87 & nan & nan & nan &  \\
10.0 & 1.5 & 1.26 & 1.28 & 1.32 & 1.36 & 1.41 & 1.52 & 1.65 & 1.86 & nan & nan & nan &  \\
10.0 & 2.0 & 1.34 & 1.36 & 1.39 & 1.43 & 1.48 & 1.57 & 1.7 & 1.89 & 2.34 & nan & nan &  \\
10.0 & 2.4 & 1.39 & 1.41 & 1.44 & 1.47 & 1.52 & 1.62 & 1.73 & 1.92 & 2.35 & 2.97 & nan &  \\
10.0 & 3.6 & 1.52 & 1.54 & 1.57 & 1.6 & 1.64 & 1.73 & 1.84 & 2.02 & 2.42 & 2.99 & 4.0 &  \\
10.0 & 5.5 & 1.67 & 1.69 & 1.71 & 1.74 & 1.79 & 1.87 & 1.98 & 2.15 & 2.56 & 3.09 & 4.0 &  \\
10.0 & 8.5 & 1.85 & 1.86 & 1.89 & 1.91 & 1.95 & 2.04 & 2.15 & 2.28 & 2.71 & 3.24 & 4.11 &  \\
10.0 & 13.0 & 2.03 & 2.04 & 2.07 & 2.09 & 2.14 & 2.22 & 2.31 & 2.4 & 2.91 & 3.43 & 4.28 &  \\
10.0 & 20.0 & 2.23 & 2.24 & 2.26 & 2.29 & 2.33 & 2.42 & 2.47 & 2.64 & 3.13 & 3.71 & 4.51 &  \\
1000.0 & 1.0 & 2.18 & 2.6 & 3.45 & 4.66 & nan & nan & nan & nan & nan & nan & nan &  \\
1000.0 & 1.5 & 1.86 & 2.05 & 2.34 & 2.65 & 3.06 & 4.01 & 5.56 & nan & nan & nan & nan &  \\
1000.0 & 2.0 & 1.78 & 1.9 & 2.1 & 2.25 & 2.49 & 2.95 & 3.54 & 4.7 & nan & nan & nan &  \\
1000.0 & 2.4 & 1.76 & 1.85 & 2.01 & 2.14 & 2.3 & 2.64 & 3.06 & 3.8 & nan & nan & nan &  \\
1000.0 & 3.6 & 1.77 & 1.83 & 1.93 & 2.02 & 2.13 & 2.32 & 2.59 & 2.99 & 4.05 & nan & nan &  \\
1000.0 & 5.5 & 1.83 & 1.88 & 1.95 & 2.01 & 2.1 & 2.23 & 2.42 & 2.7 & 3.4 & 4.48 & 6.8 &  \\
1000.0 & 8.5 & 1.94 & 1.98 & 2.03 & 2.08 & 2.14 & 2.25 & 2.41 & 2.65 & 3.18 & 3.99 & 5.45 &  \\
1000.0 & 13.0 & 2.08 & 2.11 & 2.14 & 2.17 & 2.22 & 2.34 & 2.48 & 2.68 & 3.17 & 3.84 & 5.0 &  \\
1000.0 & 20.0 & 2.21 & 2.23 & 2.27 & 2.31 & 2.36 & 2.47 & 2.6 & 2.79 & 3.25 & 3.87 & 4.89 &  \\
\enddata

%% Include any \tablenotetext{key}{text}, \tablerefs{ref list},
%% or \tablecomments{text} between the \enddata and 
%% \end{deluxetable} commands

%% General table comment marker
\tablecomments{The planet radii values in $R_\oplus$. Format and specific mass and flux values are taken to ease comparison with those from Lopez \& Fortney (2014).}

%% No \tablerefs indicated
\label{tab:LF14}
\end{deluxetable}

\begin{deluxetable}{cccccccccccccc}

%% Keep a portrait orientation

%% Over-ride the default font size
%% Use Default (12pt)

%% Use \tablewidth{?pt} to over-ride the default table width.
%% If you are unhappy with the default look at the end of the
%% *.log file to see what the default was set at before adjusting
%% this value.

%% This is the title of the table.
\tablecaption{Planetary Radii Table of Sub-Neptune Mass Planets with Ice-Rock Interiors at 10 Gyr}

%% This command over-rides LaTeX's natural table count
%% and replaces it with this number.  LaTeX will increment 
%% all other tables after this table based on this number
\tablenum{3}

%% The \tablehead gives provides the column headers.  It
%% is currently set up so that the column labels are on the
%% top line and the units surrounded by ()s are in the 
%% bottom line.  You may add more header information by writing
%% another line between these lines. For each column that requries
%% extra information be sure to include a \colhead{text} command
%% and remember to end any extra lines with \\ and include the 
%% correct number of &s.
\tablehead{\colhead{Stellar Flux} & \colhead{Planet Mass} & \colhead{0.01\%} & \colhead{0.02\%} & \colhead{0.05\%} & \colhead{0.1\%} & \colhead{0.2\%} & \colhead{0.5\%} & \colhead{1.0\%} & \colhead{2.0\%} & \colhead{5.0\%} & \colhead{10\%} & \colhead{20\%} & \colhead{} \\ 
\colhead{} & \colhead{} & \colhead{} & \colhead{} & \colhead{} & \colhead{} & \colhead{} & \colhead{} & \colhead{} & \colhead{} & \colhead{} & \colhead{} & \colhead{} & \colhead{} } 

%% All data must appear between the \startdata and \enddata commands
\startdata
0.1 & 1.0 & 1.36 & 1.38 & 1.41 & 1.45 & 1.49 & 1.58 & 1.68 & 1.83 & 2.19 & 2.62 & 3.51 &  \\
0.1 & 1.5 & 1.51 & 1.52 & 1.55 & 1.58 & 1.63 & 1.71 & 1.81 & 1.95 & 2.29 & 2.73 & 3.49 &  \\
0.1 & 2.0 & 1.62 & 1.63 & 1.66 & 1.69 & 1.73 & 1.81 & 1.91 & 2.05 & 2.37 & 2.8 & 3.53 &  \\
0.1 & 2.4 & 1.69 & 1.71 & 1.74 & 1.77 & 1.8 & 1.88 & 1.98 & 2.12 & 2.44 & 2.86 & 3.57 &  \\
0.1 & 3.6 & 1.87 & 1.89 & 1.91 & 1.94 & 1.98 & 2.05 & 2.15 & 2.29 & 2.6 & 3.01 & 3.71 &  \\
0.1 & 5.5 & 2.08 & 2.09 & 2.12 & 2.14 & 2.18 & 2.26 & 2.35 & 2.49 & 2.8 & 3.22 & 3.91 &  \\
0.1 & 8.5 & 2.31 & 2.32 & 2.35 & 2.37 & 2.41 & 2.49 & 2.58 & 2.72 & 3.03 & 3.47 & 4.15 &  \\
0.1 & 13.0 & 2.56 & 2.57 & 2.59 & 2.62 & 2.65 & 2.73 & 2.82 & 2.97 & 3.23 & 3.74 & 4.43 &  \\
0.1 & 20.0 & 2.82 & 2.83 & 2.86 & 2.88 & 2.92 & 3.0 & 3.09 & 3.2 & 3.6 & 4.04 & 4.75 &  \\
10.0 & 1.0 & 1.58 & 1.63 & 1.7 & 1.77 & 1.85 & 2.0 & 2.17 & 2.44 & nan & nan & nan &  \\
10.0 & 1.5 & 1.68 & 1.72 & 1.78 & 1.83 & 1.89 & 2.01 & 2.15 & 2.37 & nan & nan & nan &  \\
10.0 & 2.0 & 1.77 & 1.8 & 1.85 & 1.89 & 1.95 & 2.06 & 2.18 & 2.37 & 2.82 & nan & nan &  \\
10.0 & 2.4 & 1.83 & 1.86 & 1.9 & 1.94 & 2.0 & 2.1 & 2.22 & 2.4 & 2.82 & 2.97 & nan &  \\
10.0 & 3.6 & 1.99 & 2.01 & 2.04 & 2.08 & 2.13 & 2.22 & 2.33 & 2.49 & 2.87 & 3.39 & 4.0 &  \\
10.0 & 5.5 & 2.17 & 2.19 & 2.22 & 2.25 & 2.29 & 2.38 & 2.48 & 2.64 & 3.0 & 3.49 & 4.35 &  \\
10.0 & 8.5 & 2.38 & 2.4 & 2.43 & 2.46 & 2.5 & 2.58 & 2.68 & 2.83 & 3.18 & 3.66 & 4.47 &  \\
10.0 & 13.0 & 2.62 & 2.63 & 2.66 & 2.68 & 2.72 & 2.8 & 2.9 & 3.06 & 3.42 & 3.89 & 4.66 &  \\
10.0 & 20.0 & 2.87 & 2.88 & 2.9 & 2.93 & 2.97 & 3.05 & 3.15 & 3.27 & 3.68 & 4.15 & 4.93 &  \\
1000.0 & 1.0 & 3.65 & 2.6 & 3.45 & 4.66 & nan & nan & nan & nan & nan & nan & nan &  \\
1000.0 & 1.5 & 2.8 & 3.21 & 2.34 & 2.65 & 3.06 & 4.01 & 5.56 & nan & nan & nan & nan &  \\
1000.0 & 2.0 & 2.58 & 2.81 & 2.1 & 3.62 & 4.16 & 5.25 & 6.79 & 4.7 & nan & nan & nan &  \\
1000.0 & 2.4 & 2.49 & 2.67 & 2.01 & 3.24 & 3.6 & 4.25 & 5.04 & 3.8 & nan & nan & nan &  \\
1000.0 & 3.6 & 2.42 & 2.55 & 2.7 & 2.86 & 3.05 & 3.35 & 3.69 & 4.24 & 4.05 & nan & nan &  \\
1000.0 & 5.5 & 2.46 & 2.54 & 2.65 & 2.73 & 2.85 & 3.06 & 3.26 & 3.6 & 4.35 & 5.49 & 6.8 &  \\
1000.0 & 8.5 & 2.57 & 2.62 & 2.69 & 2.75 & 2.84 & 2.99 & 3.14 & 3.39 & 3.94 & 4.7 & 5.45 &  \\
1000.0 & 13.0 & 2.7 & 2.74 & 2.81 & 2.86 & 2.93 & 3.05 & 3.18 & 3.38 & 3.84 & 4.48 & 5.58 &  \\
1000.0 & 20.0 & 2.9 & 2.93 & 2.98 & 3.02 & 3.07 & 3.17 & 3.29 & 3.49 & 3.91 & 4.48 & 5.42 &  \\
\enddata

%% Include any \tablenotetext{key}{text}, \tablerefs{ref list},
%% or \tablecomments{text} between the \enddata and 
%% \end{deluxetable} commands

%% General table comment marker
\tablecomments{The planet radii values in $R_\oplus$. Format and specific mass and flux values are identical to those from Table 2.}

%% No \tablerefs indicated
\label{tab:lowMpicerock}
\end{deluxetable}

\begin{deluxetable}{ccc}
\tablecaption{Best fit coefficients to analytically estimate the radius of a planet at specified mass, composition, irradiation flux, and age using Equation~\ref{eq:Rpfit}.}
\tablenum{4}

\tablehead{\colhead{Coefficient} & \colhead{Rocky-Core Planets} & \colhead{Ice-Rock Core Planets}} 

%% All data must appear between the \startdata and \enddata commands
\startdata
$c_0$	& $	0.131	\pm	0.006	$ & $	0.169	\pm	0.008	$ \\
$c_1$	& $	-0.348	\pm	0.008	$ & $	-0.436	\pm	0.013	$ \\
$c_2$	& $	0.631	\pm	0.003	$ & $	0.572	\pm	0.005	$ \\
$c_3$	& $	0.104	\pm	0.006	$ & $	0.154	\pm	0.009	$ \\
$c_4$	& $	-0.179	\pm	0.005	$ & $	-0.173	\pm	0.007	$ \\
$c_{12}$	& $	0.028	\pm	0.002	$ & $	0.014	\pm	0.003	$ \\
$c_{13}$	& $	-0.168	\pm	0.002	$ & $	-0.210	\pm	0.003	$ \\
$c_{14}$	& $	0.008	\pm	0.003	$ & $	0.006	\pm	0.005	$ \\
$c_{23}$	& $	-0.045	\pm	0.001	$ & $	-0.048	\pm	0.001	$ \\
$c_{24}$	& $	-0.036	\pm	0.001	$ & $	-0.040	\pm	0.002	$ \\
$c_{34}$	& $	0.031	\pm	0.002	$ & $	0.031	\pm	0.002	$ \\
$c_{11}$	& $	0.209	\pm	0.005	$ & $	0.246	\pm	0.007	$ \\
$c_{22}$	& $	0.086	\pm	0.001	$ & $	0.074	\pm	0.001	$ \\
$c_{33}$	& $	0.052	\pm	0.002	$ & $	0.059	\pm	0.003	$ \\
$c_{44}$	& $	-0.009	\pm	0.002	$ & $	-0.006	\pm	0.003	$ \\
adjusted $R^2$	&	0.993			&	0.987			\\
rms error	&	0.035			&	0.047			\\
$R_{\rm core}$ & $\approx0.97M_{\rm core}^{0.28}$ & $\approx1.27M_{\rm core}^{0.27}$ 
\enddata

%% Include any \tablenotetext{key}{text}, \tablerefs{ref list},
%% or \tablecomments{text} between the \enddata and 
%% \end{deluxetable} commands

%% General table comment marker
\tablecomments{The radius contribution of the planet envelope is fit with a quadratic model,  $\log_{10}\left(\frac{R_p-R_{\rm core}}{R_{\oplus}}\right) = c_0 + \sum_{i=1}^{4} c_ix_i +\sum_{i=1}^{4}\sum_{j\geq i}^{4} c_{ij}x_ix_j$, where $x_1=\log_{10}\left(\frac{M_p}{M_{\oplus}}\right)$, $x_2=\log_{10}\left(\frac{f_{\rm env}}{0.05}\right)$, $x_3=\log_{10}\left(\frac{F_p}{F_{\oplus}}\right)$, and $x_4=\log_{10}\left(\frac{t}{5~\rm{Gyr}}\right)$. These fitting formulae are valid for $10^{-4}\leq f_{\rm env} \leq 0.2$, $1\leq M_p/M_{\oplus} \leq 20$, $4\leq F_p/F_{\oplus} \leq 400$, and $100~\rm{Myr}\leq t \leq 10~\rm{Gyr}$. 
}

%% No \tablerefs indicated
\label{tab:plaws}
\end{deluxetable}

\end{document}